\documentclass[aps,pra,superscriptaddress,footinbib,notitlepage,twocolumn]{revtex4-2}

\usepackage{tensor}
\usepackage{amsmath}
\usepackage{amsthm}
\usepackage{graphicx}
\usepackage{float}
\usepackage{comment}
\usepackage{amsfonts}
\usepackage{bm}
\usepackage{braket}
\usepackage{xcolor}
\usepackage[colorlinks=true,linkcolor=cyan,urlcolor=teal,citecolor=cyan]{hyperref}

\usepackage{enumerate}
\usepackage{mathtools}
\usepackage{bbm}
\usepackage{mathtools}
\usepackage{physics}
\usepackage{quantikz}
\usepackage{dsfont}
\usepackage{subcaption}  
\usepackage{comment}

\usepackage{siunitx}

\usepackage{makecell}

\newcommand{\ii}{\mathrm{i}}

\usepackage{quantikz}
\usepackage{mathdots}

\begin{document}

\title{Experimental demonstration that qubits can be cloned at will, if encrypted with a single-use decryption key}

\author{Koji Yamaguchi}
\thanks{These authors contributed equally to this work}
\affiliation{Department of Informatics, Faculty of Information Science and Electrical Engineering,
Kyushu University, 744 Motooka, Nishi-ku, Fukuoka, 819-0395, Japan}

\author{Leon Rullk\"otter}
\thanks{These authors contributed equally to this work}
\affiliation{Fraunhofer Institute for Industrial Engineering (IAO), Nobelstra{\ss}e 12, 70569 Stuttgart, Germany}
\author{Ibrahim Shehzad}
\affiliation{IBM Quantum, Thomas J. Watson Research Center, 1101 Kitchawan Rd, Yorktown Heights, NY 10598, USA}
\author{Sean~J.~Wagner}
\affiliation{IBM, Markham, ON L6G 1C7, Canada}
\author{Christian Tutschku}
\affiliation{Fraunhofer Institute for Industrial Engineering (IAO), Nobelstra{\ss}e 12, 70569 Stuttgart, Germany}
\author{Achim Kempf}
\affiliation{Department of Applied Mathematics, University of Waterloo, Waterloo, ON N2L 3G1, Canada}
\affiliation{Department of Physics, University of Waterloo, Waterloo, ON N2L 3G1, Canada}
\affiliation{Institute for Quantum Computing, University of Waterloo, Waterloo, ON N2L 3G1, Canada}
\affiliation{Perimeter Institute for Theoretical Physics, Waterloo, Ontario N2L 2Y5, Canada}

\begin{abstract}
The no-cloning theorem forbids the creation of identical copies of qubits, thereby imposing strong limitations on quantum technologies. 
A recently-proposed protocol, \textit{encrypted cloning}, showed, however, that the creation of perfect clones is theoretically possible - if the clones are simultaneously encrypted with a single-use decryption key. 
It has remained an open question, however, whether encrypted cloning is stable under hardware noise and thus practical as a quantum primitive.
This is nontrivial because spreading quantum information widely could dilute it until barely exceeding the noise level, leading to catastrophic fidelity decay.
Given the complexity of hardware noise, theory and classical simulation are insufficient to settle this.
Here, we settle this question experimentally, on IBM Heron-R2 superconducting processors using up to 154 qubits. 
We find that encrypted cloning is stable under hardware noise, even when used as a module, namely in parallel, series or interleaved, while preserving pre-existing entanglement. 
This establishes it as a versatile quantum primitive for practical use, and it necessitates a refinement to our understanding of the no-cloning theorem: quantum information can be spread at will, in theory and in practice, without dilution or degradation, if encrypted or obscured. The actual constraint is that the decryption mechanism must be single-use. 
\end{abstract}

\maketitle

The ability to copy data is essential in classical computing and is also routinely used, for example, for redundant cloud storage.
In contrast, the no-cloning theorem, a foundational result of quantum mechanics, establishes the impossibility of creating exact copies of quantum data~\cite{wootters_single_1982,dieks_communication_1982}. 
Since this prohibition against perfect duplication presents a significant obstacle in quantum information processing, various strategies for achieving at least some level of approximate quantum cloning have been developed~\cite{buzek_quantum_1996,gisin_optimal_1997,brus_optimal_1998,werner_optimal_1998,duan_probabilistic_1998,pati_quantum_1999,fan_quantum_2001,wang_unified_2011}.

Among these, the universal quantum cloning machine~\cite{buzek_quantum_1996} enables deterministic generation of two approximate copies of an arbitrary qubit with a fidelity of $5/6$. This value was proven to be approximate cloning's maximum~\cite{gisin_optimal_1997,werner_optimal_1998}. 
Even this maximum performance is incapable, however, to provide the clones with sufficient quantumness. For example, they cannot violate the Clauser--Horne--Shimony--Holt (CHSH) inequalities~\cite{bell_einstein_1964,clauser_proposed_1969} (see Supplementary Information).

Subsequent studies have extended approximate cloning to include schemes that exploit prior knowledge of the states to be copied, such as state-dependent~\cite{brus_optimal_1998} and probabilistic~\cite{duan_probabilistic_1998,pati_quantum_1999} cloning, as well as phase-covariant generalizations~\cite{fan_quantum_2001,wang_unified_2011} (see, e.g., Refs.~\cite{scarani_quantum_2005,fan_quantum_2014} for review). These theoretical developments have been followed by a series of experimental demonstrations across different physical platforms~\cite{lamas-linares_experimental_2002,sciarrino_realization_2005,du_experimental_2005,chen_experimental_2007,nagali_optimal_2009,pan_solid-state_2011,peng_cloning_2020,yang_experimental_2021}. While these experiments have confirmed the practical feasibility of these methods, the problem has remained that all of these methods are providing either limited fidelity or possess limited success probability or can only replicate subsets of states. 

In this context, a recent theoretical proposal, termed \emph{encrypted cloning}~\cite{yamaguchi_encrypted_2026},
has shown that, in theory, it is possible to deterministically create any number of perfect clones of an arbitrary state if, during the cloning process, the clones are encrypted with a quantum single-use decryption key (see Methods): among the encrypted clones, one can freely choose any one to decrypt and thereby recovers the original state with fidelity up to 1. The decryption process consumes the decryption key, thereby rendering all remaining encrypted clones indecipherable, which ensures consistency with the no-cloning theorem. 

For example, encrypted cloning enables, therefore, redundant quantum cloud storage: a quantum cloud storage provider hosts encrypted clones of a client's quantum data on separate servers. As long as at least one of the servers survives, the client can recover all of their data perfectly by decrypting the surviving clones. Quantum cloud storage could serve as a platform also for quantum cloud computing.  

More generally, encrypted cloning could serve as a versatile new redundancy-providing quantum primitive for quantum technologies.
We use the term quantum primitive here to denote basic quantum algorithmic building blocks (such as quantum teleportation or entanglement swapping) that can be inserted as modules, in parallel, in series or interleaved - while respecting any pre-existing entanglement of its inputs. 

It has remained an open question, however, whether encrypted cloning is stable under hardware noise and thus viable as a practical quantum primitive.
This is a nontrivial question because spreading quantum information widely could, intuitively speaking, dilute the quantum information  until it barely exceeds the noise, potentially leading to a catastrophic decay of entanglement fidelity.
If the vulnerability of encrypted cloning to noise were to grow fast enough with the number of encrypted clones, encrypted cloning could be unstable even on future error-corrected hardware.

To determine the vulnerability of encrypted cloning to hardware noise either by theoretical analysis or by classical simulation would be unreliable due to the complex nature of hardware noise. For example, perturbative reasoning, or the use of factorized noise models, would be insufficient. And even if the ideal circuits are simulatable using only Clifford gates, realistic noise generically breaks the stabilizer picture and precludes efficient classical simulation at scale. 

Therefore, we here address this question experimentally on a standard IBM Heron R2 superconducting quantum processor. 
Our experiments show that encrypted cloning is not catastrophically sensitive to hardware noise. Instead, its sensitivity to hardware noise scales particularly favorably: we find that the fidelity of encrypted cloning is 
essentially insensitive to the number of encrypted clones created. Instead, the fidelity decay is merely incremental and dominated by any quantum algorithm's minimum inevitable fidelity decay that comes with increasing circuit depth on non-error-corrected machines.

This allows us to experimentally establish that encrypted cloning can be used as a quantum primitive already on present non-error-corrected quantum hardware, i.e., as a module that can be used in series, in parallel or interleaved, while it respects pre-existing two-partite and multi-partite entanglement of its input qubits. 

To demonstrate that encrypted cloning can be used as a quantum primitive in the presence of hardware noise, i.e., as a versatile module, we performed encrypted cloning experiments that (1) measure its sensitivity to hardware noise, (2) demonstrate interleaving, (3) demonstrate operation in series and (4) demonstrate operation in parallel. These experiments also establish the preservation of pre-existing two-partite and multi-partite entanglement in the inputs. 

\medskip
\noindent{\bf Experiment~1: Measurement of sensitivity to hardware noise and demonstration that prior entanglement persists through encrypted cloning and decryption.}

\noindent
We prepare a qubit $A$ in a Bell state with an ancilla, $\tilde{A}$. Then, we encrypted-clone the qubit $A$ into $n$ encrypted clones $S_1,\ldots,S_n$. Next, we decrypt a chosen clone (we use $S_1$ without loss of generality by symmetry). 

\begin{figure}[htbp]
  \centering
  \includegraphics[width=\linewidth]{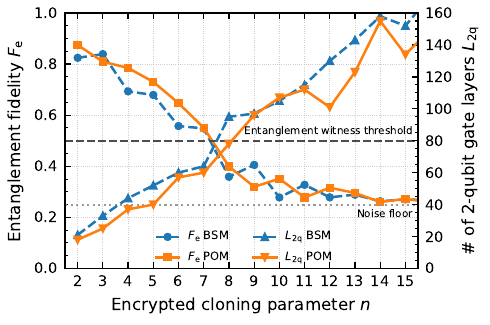} 
  \caption{Experiment~1: Proof on hardware that encrypted cloning can serve as a quantum primitive in the sense that it respects its input's pre-existing entanglement: pre-existing entanglement of qubit $A$ is measurably recovered after decrypting one of up to seven encrypted clones of $A$.    
  The drop in entanglement fidelity is dominated by the number of 2-qubit gate layers. 
  Results depend on when and which chip is used, but on a fixed chip, for a given time, the sampling errors (for 1000s of shots) are very small (see Extended Data: Table~\ref{tab:EC_fidelity}, Extended Data: Fig.~\ref{fig:comp_EM}, and Supplementary Information).
  Results confirmed with two independent measurement methods: Bell state measurement (BSM) and parity oscillations method (POM) (see Methods). They possess similar circuit depths and correspondingly similar entanglement fidelities. }
  \label{fig:gate_depth_vs_n}
\end{figure}

Then, to test whether encrypted cloning and decryption respects the prior entanglement, i.e., 
whether the decrypted clone is entangled with $\tilde{A}$, we measure its entanglement fidelity $F_e$~\cite{schumacher_sending_1996,horodecki_general_1999,nielsen_simple_2002}, i.e., its fidelity with the original Bell state, see Fig.~\ref{fig:gate_depth_vs_n}. Across $n=2$ to $n=15$, the observed fidelity decreases gradually as the circuit depth increases, rather than dropping catastrophically. In particular, entanglement is witnessed ($F_e>1/2$) up to $n=7$, and the signal remains above the maximally mixed noise floor ($F_e>1/4$) up to approximately $n\approx 13$ (Fig.~\ref{fig:gate_depth_vs_n}; Extended Data Table~\ref{tab:EC_fidelity}). The dependence of $F_e$ on hardware-induced depth overhead (two-qubit gate layers) further indicates that the dominant degradation is tracking accumulated experimental error rather than an intrinsic exponential instability with clone number.

This addresses the key screening question for modularity: even on current non-error corrected hardware, prior entanglement persists through encrypted cloning and decryption, supporting encrypted cloning as a viable candidate quantum primitive.

\medskip
\medskip
\noindent{\bf Experiment~2: Feasibility of interleaving encryption and decryption}
\begin{figure*}[thbp]
    \centering
    \begin{subfigure}[c]{0.45\textwidth}
        \centering
        \begin{minipage}[c][0.6\textwidth]{\linewidth} 
            \centering
            \begin{quantikz}[transparent,
                             row sep={0.6cm,between origins},
                             column sep = 0.4cm,
                             font=\scriptsize]
                \makeebit[-60]{$\Tilde{A},A$} & \qw & \qw\slice{} &
                \meter[style={draw={rgb,255:red,31; green,119; blue,180}}]{1}\slice{} & \qw\slice{} &
                \meter[style={draw={rgb,255:red,255; green,127; blue,14}}]{2}\slice{} &
                \meter[style={draw={rgb,255:red,44; green,160; blue,44}}]{3} \\
                & \qw & \gate[2]{\mathrm{enc}} & \qw & \qw & \qw & \qw\\
                \makeebit[-60]{$S,N$} &\qwbundle{n} & \qw & \qw &
                \gate[2]{\mathrm{dec}} & \meter{}\\
                & \qwbundle{n} & \qw & \qw & \qw & \qw & \qw
            \end{quantikz}
        \end{minipage}
        \caption{}
        \label{fig:CHSH-a}
    \end{subfigure}
    \hfill
    \begin{subfigure}[c]{0.45\textwidth}
        \centering
        \begin{minipage}[c][0.6\textwidth]{\linewidth}
            \centering
            \includegraphics[width=\linewidth]{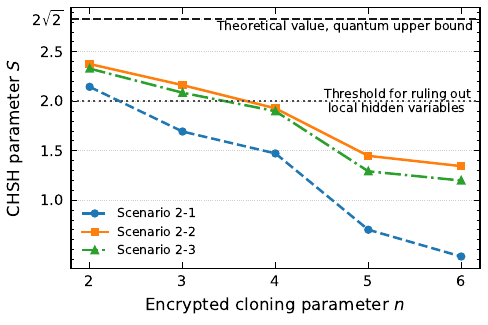}
        \end{minipage}
        \caption{}
        \label{fig:CHSH-b}
    \end{subfigure}

    \caption{Experiment 2: (a) Setup to test if the two parts of encrypted cloning, namely the creation of encrypted clones on one hand and the choosing and decrypting of one of the encrypted clones on the other can be treated, as a quantum primitive on hardware, as independent modules that can be interleaved, for example, with a measurement operation. To this end, the measurement on the ancilla $\tilde{A}$ (top rail) is performed either before, at the same time or after choosing and decrypting an encrypted clone, i.e.,   
    at time 1, 2 or 3 for experiment Scenarios 2-1, 2-2 or 2-3 respectively. Notice that this is a nontrivial test because, for example, in Scenario 2-1, the measurement on $\tilde{A}$ to test for the preservation of its pre-existing entanglement with $A$ is performed even before the choice is made which encrypted clone to decrypt and then measure.  
    (b) The results show in each scenario that encrypted cloning respects pre-existing entanglement of its input, in the sense that the CHSH inequality is violated, proving quantumness of the correlations, on the given hardware, for up to three encrypted clones.
The deterioration of the performance is dominated by the duration of the circuit. For example, the lower performance in Scenario 2-1 arises from the extra idling time during the measurement of $\tilde{A}$. 
The statistical error bars from $10,000$ shots are very small (see also Extended Data: Table~\ref{tab:S_vs_n}).} 
    \label{fig:CHSH}
\end{figure*}

\noindent
For encrypted cloning to be able to serve as a candidate quantum primitive, its two parts (encrypted cloning and decryption) should, in practice, be reliably usable as modules that can be separated in time and interleaved with other operations, such as measurements, while still respecting the prior entanglement of its input registers. 

To test this, we again start from a Bell pair of qubits $A$ and $\tilde{A}$, apply encrypted cloning to $A$, and then implement three distinct measurement orderings within one circuit (Fig.~\ref{fig:CHSH-a}). In Scenario 2-1, we measure $\tilde{A}$ first and decrypt and measure a clone later; in Scenario 2-2, we decrypt and then measure $\tilde{A}$ and the decrypted clone at the same time; in Scenario 2-3, we decrypt and measure the clone first and measure $\tilde{A}$ later. These orderings probe whether intermediate measurements and idling between the two stages destroy the recovered nonlocal correlations.

We quantify recovered nonclassicality by the CHSH parameter $S$~\cite{bell_einstein_1964,clauser_proposed_1969} between $\tilde{A}$ and the decrypted clone (Fig.~\ref{fig:CHSH-b}; Extended Data Table~\ref{tab:S_vs_n}). We observe a violation $|S|>2$ for up to $n=3$ (corresponding to up to three encrypted clones), demonstrating that encryption followed by later decryption can restore correlations strong enough to rule out local hidden-variable models in the accessible regime. 

These tests demonstrate that encrypted cloning can be used in a modular fashion in the sense that encrypted cloning and decryption can be separated in time and even interleaved with measurements. In particular, the measurement of $\tilde{A}$ can be chosen to occur before, after, or simultaneously (and therefore, in principle, also spacelike separated) from the choice of which clone is decrypted, i.e., encrypted cloning is compatible with delayed choices of measurement or decryption.

These results show that separating and interleaving encrypted cloning and its decryption are feasible in practice, supporting encrypted cloning as a flexible quantum primitive.

\medskip
\noindent{\bf Experiment~3: Operation in series: Feasibility of iterating encrypted cloning}
\newline
\noindent
A further criterion for a quantum operation to be a usable primitive is composability with itself. In theory, encrypted cloning admits an iterated construction~\cite{yamaguchi_encrypted_2026} in which the number of encrypted clones grows exponentially, while the decryption key required to recover a chosen final clone grows only linearly with the number of iterations. In practice, in the presence of noise, iterating encrypted cloning provides a stringent stress test of operational stability, because iteration increases both circuit depth and the size of the multipartite entangled state.

We implement this iterated scheme on hardware by starting again with a Bell pair, $A,\tilde{A}$. We then create three encrypted clones of $A$ and of the all subsequent encrypted clones, iterating $l$ times. Eventually, after decrypting a selected final clone, we measure its entanglement fidelity $F_e$  with $\tilde{A}$ (Extended Data Table~\ref{tab:iterated_fidelity}). With maximal chip use (154 qubits), generating 77 encrypted clones, we still observe recoverability above the noise floor ($F_e>1/4$). Moreover, we retain an entanglement witness ($F_e>1/2$) for up to 27 encrypted clones.

These results support treating encrypted cloning as a composable quantum primitive since they show that iteration of encrypted cloning does not induce an abrupt loss of recoverability. To the contrary, we find that since, by iteration, the number of encrypted clones grows exponentially with the circuit depth, iterated encrypted cloning is in practice preferable over large $n$ direct encrypted cloning. 
The much shorter key length and shallower circuit depth of iterated encrypted cloning allow the creation of more encrypted clones for any desired fidelity than through large $n$ one-off encrypted cloning on the same processor.

\medskip
\noindent{\bf Experiment~4: Operation in parallel: Feasibility of encrypted cloning inside multipartite circuits}
\newline
\noindent
In practice, in quantum algorithms, qubits participate in larger entangled registers. For encrypted cloning to be useful as a practical quantum primitive it should, therefore, preserve the embedding of its input qubit in a multi-partite entangled structure. The operational requirement is that decrypting an encrypted clone of $A$ restores $A$'s prior correlations with the rest of the system, even though the encryption stage had spread information across many encrypted clones. In addition, as a practical quantum primitive, encrypted cloning should be applicable in parallel to multiple qubits.

\begin{figure}[htbp]
  \centering
  \includegraphics[width=\linewidth]{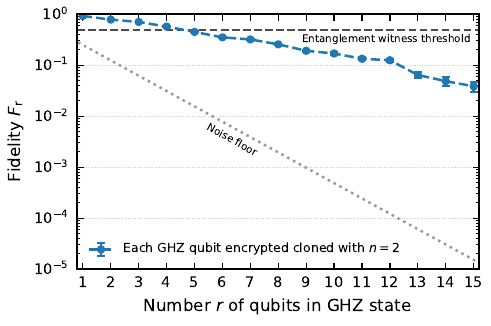} 
  \caption{Experiment 4: 
  We prepare GHZ states for $r$ up to $15$. We then independently create $3$ encrypted clones of each of the $r$ qubits, i.e., we produce $r$ independent groups of three encrypted clones. After the encrypted cloning, in each group, one of the clones is decrypted. Finally, the fidelity $F_r$ between the recovered GHZ state and the initial GHZ state is measured using the POM with 10 000 shots per measurement setting. The signal always remains well above the noise floor and reaches fidelities above the entanglement witness threshold for GHZ states with up to 4 qubits. 
  These results support that encrypted cloning can serve as a candidate quantum primitive in the sense that it respects not only bipartite but also genuine multipartite entanglement of input qubits embedded in larger circuits, and encrypted cloning is a quantum primitive in the sense that it can also be applied in true modular fashion in parallel to multiple qubits.}
  \label{fig:gate_depth_vs_r}
\end{figure}

We test this strongest modularity criterion using $r$-qubit Greenberger--Horne--Zeilinger (GHZ) states~\cite{greenberger_bells_1990}. We prepare GHZ states for $r$ up to 15, then independently create three encrypted clones of each GHZ qubit, producing $r$ groups of three encrypted clones and keys. We subsequently decrypt one clone from each group. The challenge is to determine whether these decrypted clones are then in the original GHZ state. To this end, we measure the fidelity $F_r$ of the recovered $r$-qubit state with the original GHZ state (Fig.~\ref{fig:gate_depth_vs_r}; Extended Data Table~\ref{tab:ghz_fidelity}). Across the tested range, the recovered GHZ fidelity remains well above the maximally mixed noise floor $2^{-r}$, and genuine multipartite entanglement is witnessed ($F_r>1/2$) up to $r=4$.

This experiment demonstrates two aspects of modularity at once. First, encrypted cloning can be applied independently and in parallel to multiple qubits such as those in a GHZ state. Second, decrypting each selected clone restores the multipartite correlation structure of the qubits' original multi-partite state, showing that encrypted cloning respects the operational role of a qubit within an entangled register. This supports viewing encrypted cloning as a quantum primitive that provides experimentally feasible modules that can be inserted into multipartite algorithms: after decryption, the qubits return to the algorithm with their intended global correlations intact.

A common finding of Experiments 1--4 is that the degradation of the fidelity of encrypted cloning is merely incremental and is governed 
not by the number of encrypted clones but
by the circuit depth, which is the bare minimum degradation on non-error-corrected hardware. 
This implies that the fidelity of encrypted cloning scales exceedingly well with respect to the number of encrypted clones created on the non-error corrected hardware. This is because the number of encrypted clones can be made to grow exponentially with the circuit depth through iterated encrypted cloning as, e.g., in Experiment 3. This reflects the fact that quantum information does not become diluted when encrypted cloned. 

\medskip

\noindent {\bf Conclusion}\newline 
\noindent
Collectively, these four experiments establish that encrypted cloning is favorably robust to hardware noise in the sense that even already on current non-error-corrected processors, encrypted cloning can be used as a module, i.e., in parallel, in series or interleaved, while measurably preserving pre-existing entanglement of its input qubits.
This establishes encrypted cloning as a versatile quantum primitive for practical use in quantum technologies, including quantum infrastructure,
for example, to enable parallel quantum storage and, potentially, also parallel quantum computing. 

The experiments also necessitate a refinement to our understanding of the no-cloning theorem: quantum information can be spread at will, in theory and in practice, without dilution or degradation, namely if encrypted or obscured. The actual constraint is that the decryption mechanism must be single-use.

\medskip
\noindent
{\bf \large Acknowledgments}
\noindent

\noindent
K.Y. acknowledges support from JSPS Grants-in-Aid for Scientific Research No.\ JP24KJ0085.
A.K. acknowledges support through the Dieter Schwarz Foundation, a Discovery Grant of the National Science and Engineering Council of Canada (NSERC), an Applied Quantum Computing Challenge Grant from the National Research Council of Canada (NRC), a Discovery Project grant of the Australian Research Council (ARC) and support from Perimeter Institute which is supported in part by the Government of Canada through the Department of Innovation, Science and Industry Canada and by the Province of Ontario through the Ministry of Colleges and Universities. The contribution of L.R. and C. K. has been supported by the Fraunhofer Heilbronn Research and Innovation Centers HNFIZ.

\medskip
\noindent{\bf \large Data availability}\newline 
\noindent
The data that support the findings of this study are available in the Supplementary Information and from the authors upon reasonable request.

\medskip
\noindent{\bf \large Code availability}\newline 
\noindent
The codes used to generate the results of this study are available from the authors upon reasonable request.

\medskip
\noindent{\bf \large Author Contributions}\newline 
\noindent
K.Y., L.R., and A.K. designed the experiments and analyzed data;
L.R. developed the code and implemented the experiments;
I.S. and S.J.W. provided technical guidance on the implementation on the quantum processor;
C.T. contributed to early-stage discussions on the study design; 
K.Y. and A.K. wrote Main Body, Methods, and Supplementary Information;
L.R. and I.S. contributed Supplementary Information.

\medskip
\noindent{\bf \large Competing interests}\newline 
\noindent
The authors declare no competing interests.


%

\clearpage
\noindent
{\bf \Large Methods}

\medskip
\noindent
{\bf Encrypted cloning protocol}

\noindent
The encrypted cloning protocol~\cite{yamaguchi_encrypted_2026} provides a unitary method for creating encrypted copies of an arbitrary qubit $A$. The process is divided into two stages: the creation of encrypted clones and the decryption of an encrypted clone, which recovers the original state.

We start by preparing $n$ qubits $S_i,~i=1,\ldots,n$, called signal qubits, and $n$ qubits $N_i,~i=1,\ldots,n$, called noise qubits with each pair $(S_i,N_i)$, initialized in the maximally entangled state
\begin{align}
    \ket{\phi}_{S_iN_i}\coloneqq \frac{1}{\sqrt{2}}\left(\ket{0}_{S_i}\ket{0}_{N_i}+\ket{1}_{S_i}\ket{1}_{N_i}\right)\label{eq:bell_state_SiNi}
\end{align}
We then set aside and isolate the noise qubits $\{N_i\}$. Through their entanglement with the $\{S_i\}$, the role of the noise qubits $\{N_i\}$ is to provide the noise to the signal qubits which will be used for encryption. The fact that the noise qubits keep a record of this quantum noise will later allow the noise qubits to serve as the decryption key. 

Next, in order to turn the signal qubits $\{S_i\}$ into $n$ \it encrypted clones \rm of a qubit $A$ that is in an unknown state, we let $A$ interact with the signal qubits $\{S_i\}$ through a suitable unitary $U^{(n)}_{\mathrm{enc}}$. Since we have set aside the noise qubits, the unitary $U^{(n)}_{\mathrm{enc}}$ acts as the identity on the Hilbert space of the noise qubits, i.e., the noise qubits $\{N_i\}$ remain unaffected during encrypted cloning. Therefore, the noise qubits can carry neither classical nor quantum information about $A$. 

Later, any one of the encrypted clones $\{S_i\}$, which we denote by $S_j$, can then be decrypted to reproduce the original state of qubit $A$. The decryption, or ``denoising," of $S_j$ is accomplished by a unitary $U^{(n)}_{\mathrm{dec}}$ that acts nontrivially only on $S_j$ and on the noise qubits $\{N_i\}$, thereby recovering the original state of $A$ on $S_j$.

The encryption is performed by applying a unitary operator $U_{\mathrm{enc}}^{(n)}$ that acts on qubit $A$ and all $n$ signal qubits. This operator is given by:
\begin{align}
    U_{\mathrm{enc}}^{(n)}\coloneqq e^{-\frac{\pi\ii }{4} \sigma_{1}^{(A)}\otimes \left(\bigotimes_{i=1}^n\sigma_{1}^{(S_i)}\right)}e^{-\frac{\pi\ii}{4} \sigma_{3}^{(A)}\otimes \left(\bigotimes_{i=1}^n\sigma_{3}^{(S_i)}\right)}.\label{eq:encoding_operation}
\end{align}
Here, the superscript of each Pauli operator denotes the subsystem that this Pauli operator acts on. The decryption is performed by applying a unitary operator $U_{\mathrm{dec}}^{(n)}$ that acts on an arbitrarily chosen clone $S_j$ and the \textit{entire} set of noise qubits $\{N_i\}_{i=1}^n$, which constitute the key. Specifically, the decryption unitary is given by
\begin{align}
    U_{\mathrm{dec}}^{(n)}\coloneqq \sum_{\mu=0}^3\alpha_\mu\left(\ket{\phi_\mu}\bra{\phi_\mu}_{S_jN_j}\right)\otimes \left(\bigotimes_{\substack{i=1\\i\neq j}}^n\sigma_{\mu}^{(N_i)\top}\right) \label{eq:decoding_op},
\end{align}
where $\ket{\phi_\mu}_{S_iN_i}\coloneqq \sigma_{\mu}^{(S_i)}\otimes \mathbb{I}^{(N_i)}\ket{\phi}_{S_iN_i}$, $\top$ denotes the transpose with respect to the computational basis $\{\ket{0},\ket{1}\}$, and the coefficients are defined by $\alpha_0\coloneqq 1$, $\alpha_1=\alpha_3\coloneqq\ii$ and $\alpha_2\coloneqq -\ii^{n+1}$. 

As was shown in \cite{yamaguchi_encrypted_2026}, the properties of $U_{\mathrm{enc}}^{(n)}$ and $U_{\mathrm{dec}}^{(n)}$ justify the specific terminology \it encrypted cloning: \rm  (a) each encrypted clone is encrypted in the sense that, on its own, it is in the maximally mixed state and reveals no information about the initial state of $A$ and (b) each encrypted clone is a clone in the sense that, by decrypting it, the original unknown state of $A$ can be recovered deterministically and with perfect fidelity.

\medskip\noindent
{\bf Gate decomposition}

\begin{figure*}[t]
  \centering 
  \begin{quantikz}
        \lstick{$A$}& \ctrl{1} & \qw &\qw & \qw& \qw & \qw & \qw & \qw &\qw&\qw &\qw &\qw & \ctrl{1} & \\
        \lstick{$S_1$}& \targ{} & \ctrl{1} &\qw & \qw & \qw & \qw & \qw &\qw&\qw &\qw &\qw & \ctrl{1}& \targ{}& \\
        \lstick{$S_2$}& & \targ{}  &\wire[l][1]["\ddots"{below,pos=0.1}]{a} &\qw &\qw &\qw  &\qw &\qw& & & \wire[l][1]["\iddots"{below,pos=0.1}]{a} & \targ{}&\qw & \\
        \lstick{$S_{m-2}$} &  &  & & \ctrl{1} & &  &  &  & &\ctrl{1}& & & & \\
        \lstick{$S_{m-1}$}& \qw & \qw & \qw& \targ{} & \ctrl{1}& & & \qw & \ctrl{1} &  \targ{} &\qw &\qw&\qw& \\
        \lstick{$S_{m}$} & \qw & \qw & \qw&  & \targ{} &\ctrl{1} & &\ctrl{1}  & \targ{} &  \qw &\qw &\qw&\qw& \\
        \lstick{$S_{m+1}$}& & \qw & \qw & & \targ{} &\targ{} & \gate{R_z\left(\frac{\pi}{2}\right)} & \targ{} & \targ{}&  \qw  &\qw&\qw& & \\
        \lstick{$S_{m+2}$}& & \qw &  & \wire[l][1]["\iddots"{below,pos=0.1}]{a}&\ctrl{-1} & \qw & \qw & \qw &\ctrl{-1} & \wire[l][1]["\ddots"{below,pos=0.1}]{a}  & & & &\\
        \lstick{$S_{2m-2}$} & & \qw & \targ{} & & & \qw & \qw & \qw & &  &\targ{}& \qw& &\\
        \lstick{$S_{2m-1}$}& & \targ{}& \ctrl{-1}& & & & & & & & \ctrl{-1}& \targ{}& &\\
        \lstick{$S_{2m}$}& & \ctrl{-1} & & & & & & & & & & \ctrl{-1}& &\\
    \end{quantikz}
    \caption{Gate decomposition of first exponential factor of Eq.~\eqref{eq:encoding_operation}, i.e., $e^{-\ii \frac{\pi}{4}\sigma_{3}^{(A)}\otimes \left(\bigotimes_{i=1}^n\sigma_{3}^{(S_i)}\right)}$ when $n=2m$ with $m\in\mathbb{Z}_{>0}$. When $n=2m+1$, the encrypted cloning operation requires two additional CNOT gates applied to $S_{2m+1}$ and $S_{2m}$. 
    The second exponential factor, $e^{-\ii \frac{\pi}{4} \sigma_{1}^{(A)}\otimes \left(\bigotimes_{i=1}^n\sigma_{1}^{(S_i)}\right)}$, can be implemented in the similar way, since $e^{-\ii \frac{\pi}{4} \sigma_{1}^{(A)}\otimes \left(\bigotimes_{i=1}^n\sigma_{1}^{(S_i)}\right)}=H^{\otimes (n+1)}e^{-\ii \frac{\pi}{4} \sigma_{3}^{(A)}\otimes \left(\bigotimes_{i=1}^n\sigma_{3}^{(S_i)}\right)}H^{\otimes (n+1)}$.}\label{fig:ED_encryption_op}
\end{figure*}
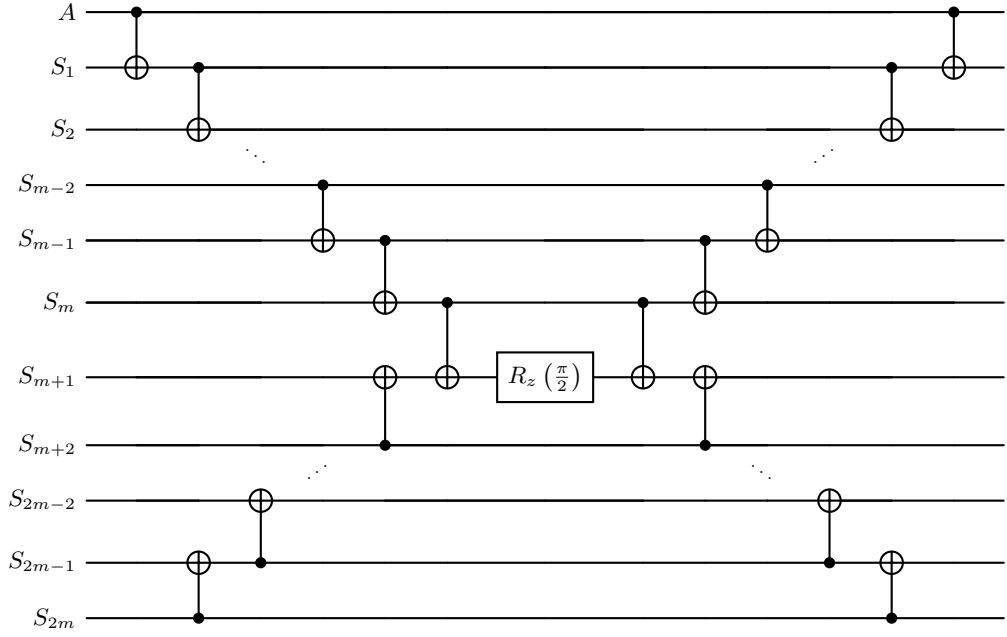

\noindent
A gate decomposition of the encryption unitary operation (Eq.~\eqref{eq:encoding_operation}) is provided in Extended Data Figure~\ref{fig:ED_encryption_op}. In this figure, each CNOT gate can be implemented by using the $\texttt{cz}$ and the Hadamard gate $H$, which can be implemented by using $\texttt{rz}$ and $\texttt{sx}$ as $H=R_z\left(\frac{\pi}{2}\right)\cdot \sqrt{X}\cdot R_z\left(\pi\right)$ (up to irrelevant global phase), where $R_z(t)\coloneqq e^{-\ii\theta \frac{\theta}{2}\sigma_z}$. 

\begin{figure*}[t]
  \centering 
  \begin{quantikz}
        \\
       \lstick{$S_1$}  & \ctrl{5}& & \ \ldots\ &    
        \ctrl{4}  &       &\ctrl{3} &        &        & \ctrl{2}& \gate{S} &\\
       \lstick{$N_1$} &         &\ctrl{4} & \ \ldots\ &  &\ctrl{3} &        &\ctrl{2} &\ctrl{1} &         & \gate{S}&\\
       \lstick{$N_2$} &        &         & \ \ldots\ &&    
          &         &          &\targ{}& \gate{Z} &         &\\
       \lstick{$N_3$}&         & & \ \ldots\ &   
                &  &\gate{Z} & \targ{}&         &          &          &\\
       \lstick{$N_4$}  & & &\wire[l][1]["\iddots"{below,pos=0.1}]{a} \ \ldots\ &\gate{Z}  &\targ{}  & &        &         &         &         &\\
       \lstick{$N_n$} &\gate{Z}& \targ{} & \ \ldots\ & &        &          &        &          &         &         &\\
    \end{quantikz}.
    \caption{Gate decomposition of $(V\otimes \mathbb{I}_{N_2\cdots N_n})U_{\mathrm{dec}}^{(n)} (V\otimes \mathbb{I}_{N_2\cdots N_n})^\dag$, where the decryption operation $U_{\mathrm{dec}}^{(n)}$ (for $S_1$) and untiary operator $V$ are given in Eqs.~\eqref{eq:decoding_op} and \eqref{eq:V_operator_basis_change}, respectively. Here, $S$ denotes the phase gate, realized as $S=R_z\left(\frac{\pi}{2}\right)$ up to a global phase.}\label{fig:ED_decryption_op}
\end{figure*}
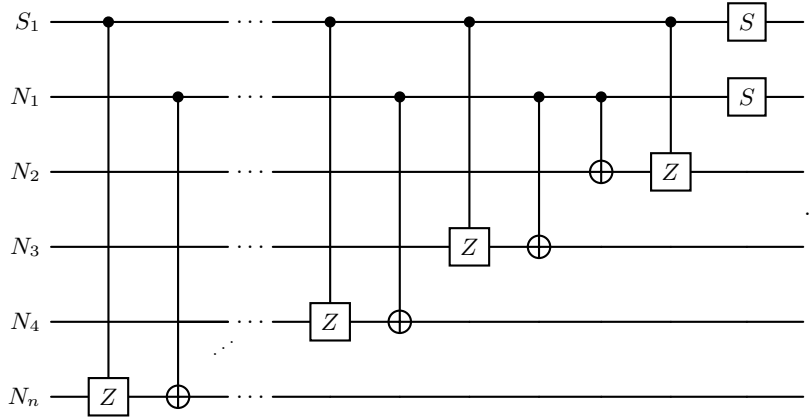

To decompose the decryption operation (Eq.~\eqref{eq:decoding_op}), we introduce a unitary operator
\begin{align}
    V&\coloneqq (H\otimes H)CZ(I\otimes H)\label{eq:V_operator_basis_change},
\end{align}
satisfying $V\ket{\phi_\mu}=\ket{\mu_1}\ket{\mu_1\oplus\mu_2}$
for $\mu=0,1,2,3$, where and $\mu=\mu_1\mu_2$ is the binary representation of $\mu$, and $\oplus$ denotes the addition modulo two. A gate decomposition of $(V\otimes \mathbb{I}_{N_2\cdots N_n})U_{\mathrm{dec}}^{(n)} (V\otimes \mathbb{I}_{N_2\cdots N_n})^\dag$ is shown in Extended Data Figure~\ref{fig:ED_decryption_op}.

\medskip
\noindent
{\bf Experimental implementation}

\noindent
All experiments in this work were performed on an IBM Heron R2 generation superconducting quantum processor, specifically the \texttt{ibm\_kingston} backend, accessed through the IBM Quantum Platform service. In Supplementary Information, we also made a comparison to the performance of the \texttt{ibm\_aachen} Heron R2 processor for Experiment 1.
IBM Heron R2 processors feature 156 physical fixed-frequency transmon qubits coupled via tunable couplers arranged in a heavy-hex topology. 
Representative values at the time of experiments for a measured median relaxation time ($T_1$) and median dephasing time ($T_2$) of qubits on the \texttt{ibm\_kingston} are 263 µs and 149 µs, respectively.

The quantum circuits used to implement the protocol were first designed at the logical level (given by Extended Data Figures~\ref{fig:ED_encryption_op} and \ref{fig:ED_decryption_op})
and then transpiled to the native gate set of the IBM Heron processor, \{\texttt{rz}, \texttt{sx}, \texttt{x}, \texttt{id}, \texttt{cz}\}, using the transpiler from Qiskit \cite{qiskit2024} with \texttt{optimization level~3}. Transpilation was performed to account for the gate directionality, and the heavy-hex topology of the \texttt{ibm\_kingston} backend. The final gate counts and circuit depths of the transpiled circuits are reported in Extended Data.
Limited qubit connectivity on the Heron processor requires use of SWAP gates in the transpiled circuits, hence reported gate counts and circuit depths for each experiment include SWAP operations implemented in the native gate set by the transpiler.

\medskip
\noindent
{\bf Fidelity estimation}

\noindent
We employ two methods to evaluate the performance of the encrypted cloning protocol via fidelity estimation of Bell states and $r$-qubit GHZ states. The fidelity of some experimental state $\rho_\mathrm{exp}$ in relation to a desired state $\ket{\phi}$ is defined as
\begin{equation}
    F_\phi = \expval{\rho_\mathrm{exp}}{\phi} \: .
\end{equation}

Firstly, the fidelity of a Bell state is measured by performing a Bell state measurement (BSM) using the following circuit
\begin{center}
    \begin{quantikz}
        \lstick{$\Tilde{A}$} & \ctrl{1} & \gate{H} & \meter{}\\
        \lstick{$A$} & \targ{} & \qw & \meter{}
    \end{quantikz}
\end{center}
and measuring both qubits in the $\sigma_z$-basis. The fidelity of the two-qubit decrypted state $\rho_\mathrm{exp}$ of $A$ and $\Tilde{A}$ with respect to the Bell state $\ket{\phi} = (\ket{00} + \ket{11})/\sqrt{2}$ can be estimated through sampling, given by
\begin{equation}
    F = \frac{N_\mathrm{00}}{N_\mathrm{shots}} \: ,
\end{equation}
where $N_\mathrm{00}$ is the number of times the outcome \texttt{00} is registered and $N_\mathrm{shots}$ is the total number of experimental samples. Error bars are calculated using the experimental shot noise. The standard deviation $\sigma_F$ of the fidelity $F$ is given by
\begin{equation}
    \sigma_F = \sqrt{\mathrm{Var}(F)} = \sqrt{\frac{F(F-1)}{N_\mathrm{shots}}}\label{eq:VarF}
\end{equation}
since $F$ is the probability of measuring the outcome 00.

An alternative approach developed in \cite{toth_detecting_2005,guhne_toolbox_2007} is the parity oscillations method (POM). The POM can be advantageous because it requires only local measurements and therefore requires fewer swaps to move qubits.
The goal is to determine the fidelity between a state $\rho$ with the $r$-partite GHZ state:
\begin{equation}
    F_r = \langle \mathrm{GHZ}_r|\rho |\mathrm{GHZ}_r\rangle=\mathrm{Tr}(\rho |\mathrm{GHZ}_r\rangle\langle \mathrm{GHZ_r}|),
\end{equation}
where $|\mathrm{GHZ}_r\rangle\langle \mathrm{GHZ}_r|$ can be decomposed into $2\dyad{\mathrm{GHZ}_r} = \dyad{0}^{\otimes r} + \dyad{1}^{\otimes r} + \dyad{0}{1}^{\otimes r} + \dyad{1}{0}^{\otimes r}$. The first two terms $P = \dyad{0}^{\otimes r} + \dyad{1}^{\otimes r}$ are measured by sampling the circuit using the measurement setting $\sigma_z^{\otimes r}$ and assigning the terms to the probabilities of the measurement outcomes $0^{\otimes r}$ and $1^{\otimes r}$. Subsequently, the $r$ measurement settings
\begin{equation}
    \mathcal{M}_k = \left [ \cos \left ( \frac{k\pi}{r} \right ) \sigma_x + \sin \left ( \frac{k\pi}{r} \right ) \sigma_y \right ]^{\otimes r}, k = 1,...,r
\end{equation}
are used to measure $\dyad{0}{1}^{\otimes r} + \dyad{1}{0}^{\otimes r}$ using the relation
\begin{equation}
    \chi = \dyad{0}{1}^{\otimes r} + \dyad{1}{0}^{\otimes r} = \frac{1}{r} \sum_{k=1}^r (-1)^k \mathcal{M}_k \: .
\end{equation}
In the experiment, we measure $\expval{\mathcal{M}_k}$ by rotating into the correct basis using an $R_z$-rotation with angle $\theta = k\pi/r$, a subsequent Hadamard gate and measuring $E_k = \expval{\sigma_z^{\otimes r}}$ in the $\sigma_z$ setting.
This amounts to 
\begin{equation}
    F_r = \frac{1}{2}(\expval{P} + \expval{\chi}) \: .
\end{equation}
The POM fidelity estimation protocol for the case of $r$-qubit GHZ states requires in total $r+1$ measurement settings.
Errors are a combination of sampling in the $\sigma_\mathrm{z}$-basis and expectation values in the $x$-$y$-plane. The variance of $P$ is calculated using the same method as in equation \eqref{eq:VarF} and amounts to
\begin{equation}
    \mathrm{Var}(P) = \frac{p_\mathrm{00}(p_\mathrm{00}-1)}{N_\mathrm{shots}} + \frac{p_\mathrm{11}(p_\mathrm{11}-1)}{N_\mathrm{shots}} \: ,
\end{equation}
where $p_\mathrm{00}$ and $p_\mathrm{11}$ are the probabilities of measuring \texttt{00} and \texttt{11} in the $\sigma_z$-basis measurement. On the other hand
\begin{equation}
    \mathrm{Var}(\chi) = \frac{1}{r^2}\sum_{k=1}^r\mathrm{Var}(E_k) = \frac{1}{r^2N_\mathrm{shots}}\sum_{k=1}^r (1-E_k^2) \: ,
\end{equation}
since the $E_k$ is derived from Bernoulli trials. Finally,
\begin{equation}
    \sigma_F = \sqrt{\mathrm{Var}(P) + \mathrm{Var}(\chi)} \: ,
\end{equation}
provides the standard deviation of the fidelity. 

\medskip
\noindent
{\bf Experiment~1: Measurement of sensitivity to hardware noise and demonstration that prior entanglement persists through encrypted cloning and decryption.}

\noindent
The first experiment implements the original encrypted cloning protocol. To quantify the output quality after cloning and decryption, we entangle the input qubit with an ancilla in a Bell state and observe the entanglement fidelity $F_e$ relative to that state. This metric is directly related to the channel's average fidelity~\cite{schumacher_sending_1996,horodecki_general_1999,nielsen_simple_2002}.

We begin by preparing a qubit $A$ and an ancilla $\tilde{A}$ in the Bell state $\ket{\phi}=(\ket{00}+\ket{11})/\sqrt{2}$. In addition, we initialize $n$ pairs of signal and noise qubits $\{(S_i, N_i)\}_i$ in the Bell state defined in Eq.~\eqref{eq:bell_state_SiNi}, where $n = 2\text{--}15$. We then apply the encryption unitary $U_{\mathrm{enc}}^{(n)}$ in Eq.~\eqref{eq:encoding_operation} to $A S_1 \ldots S_n$, thereby transforming the signal qubits $\{S_i\}_i$ into $n$ encrypted clones of $A$.

Subsequently, we decrypt one encrypted clone, $S_1$, using the full set of noise qubits $\{N_i\}_i$ by applying the decryption unitary $U_{\mathrm{dec}}^{(n)}$ specified in Eq.~\eqref{eq:decoding_op}. Then $F_e$ is measured by using the BSM and the POM.

Due to the symmetry of $U_{\mathrm{enc}}^{(n)}$, all signal qubits are equivalent. Therefore, experimentally successful decryption of $S_1$ certifies the decryptability of any $S_j$ for $j = 1, \ldots, n$.

\medskip
\noindent
{\bf Experiment~2: Feasibility of interleaving encryption and decryption}

\noindent
In this experiment, we again perform encrypted cloning of qubit $A$ that is initially maximally entangled with $\tilde{A}$, for varying $n$, but we now explore the effect of the relative timing of the measurements on $\tilde{A}$ and a decrypted clone of $A$, and we use a new figure of merit, to determine the quantumness of the result, namely the amount of violation of the CHSH inequality. 

For the motivation, let us consider that when a qubit $A$, initially maximally entangled with an ancilla $\tilde{A}$, undergoes encrypted cloning, the ancilla $\tilde{A}$ acquires multiple potential Bell partners---namely, the systems that are maximally entangled with $\tilde{A}$. In particular, each set of one $S_j$ and all noise qubits is maximally entangled with $\tilde{A}$. 
The subsequent decryption of any encrypted clone $S_j$ renders $S_j$ the sole Bell pair partner of $\tilde{A}$. Since a measurement on one subsystem can, in principle, induce state collapse on its Bell state partner, this structure allows us to demonstrate a delayed-choice form of state collapse within the encrypted-cloning framework. 

Specifically, we perform an experiment demonstrating a violation of the CHSH inequality under three distinct measurement orderings within a single quantum circuit (see Fig.~\ref{fig:CHSH-a}). 
In Scenario~2-1, after encrypted cloning of $A$, we first measure $\tilde{A}$ and then decrypt and measure an encrypted clone $S_1$. The measurement of $\tilde{A}$ induces state collapse on its entangled partner; however, which of the encrypted clones $S_j$ ultimately serves as that partner is determined only later, when $S_1$ is decrypted. 
In Scenario~2-2, after decrypting $S_1$, we measure $\tilde{A}$ and $S_1$ simultaneously.
In Scenario~2-3, the order is reversed compared to Scenario~2-1: we first decrypt and measure $S_1$, and subsequently measure $\tilde{A}$. In this case, the decryption operation determines which encrypted clone can induce state collapse on $\tilde{A}$. 

As a quantifier for identifying violations of the CHSH inequality, and therefore proof of quantumness, we adopt the standard CHSH parameter $S$ between $\tilde{A}$ and $S_1$, defined as
\begin{align}
S &\coloneqq \Braket{\sigma_3^{(\tilde{A})}\otimes \frac{\sigma_3^{(S_1)}+\sigma_1^{(S_1)}}{\sqrt{2}}}+ \Braket{\sigma_3^{(\tilde{A})}\otimes \frac{\sigma_3^{(S_1)}-\sigma_1^{(S_1)}}{\sqrt{2}}}\nonumber\\
&+\Braket{\sigma_1^{(\tilde{A})}\otimes \frac{\sigma_3^{(S_1)}+\sigma_1^{(S_1)}}{\sqrt{2}}} -\Braket{\sigma_1^{(\tilde{A})}\otimes \frac{\sigma_3^{(S_1)}-\sigma_1^{(S_1)}}{\sqrt{2}}}.
\end{align}
Here, $\langle O \rangle$ denotes the expectation value of an observable $O$ evaluated on the state after decrypting $S_1$. In theory, the decrypted clone fully recovers the entanglement with $\tilde{A}$, and the CHSH parameter attains its maximal value $2\sqrt{2}$. As is well known, a value of $|S|>2$ rules out any description based on local hidden-variable theories~\cite{clauser_proposed_1969}.

Figure~\ref{fig:CHSH-b} shows that on the given hardware, the violation of the CHSH inequalities is visible for up to three encrypted clones. It also compares the CHSH parameters for each of the relative timings of the measurements of the ancilla $\tilde{A}$ and a decrypted clone of $A$. In theory, the timing has no influence on the resulting CHSH violation. In the experiments, we find that in Scenarios 2-2 and 2-3, where only qubit $\tilde{A}$ idles while the signal and noise qubits are processed, the performance is better than in Scenario (2-1). In Scenario 2-1, all the encrypted clones and noise qubits idle, and accumulate noise, while qubit $\tilde{A}$ is being measured. This idling time, i.e., the time to measure qubit $\tilde{A}$, is about \SI{3}{\micro\second}, which is about three times the duration, about \SI{1}{\micro\second}, of the full circuit that decrypts an encrypted clone.  

We also experimentally simultaneously measured $\tilde{A}$ and an undecrypted encrypted clone $S_2$ ($n=2$) and obtained $S=0.014 \pm 0.020$, which is consistent with the theoretical expectation and the simulated result of $S=0$. This demonstrates that encrypted clones (before decryption) are uncorrelated with qubit $\tilde{A}$, which justifies the terminology ``encrypted" clone. 

\medskip
\noindent
{\bf Experiment~3: Operation in series: Feasibility of iterating encrypted cloning}

\noindent
For the encrypted cloning protocol with even $n$, the original qubit $A$ can also serve as an encrypted clone~\cite{yamaguchi_encrypted_2026}, producing a total of $n+1$ encrypted clones. By iterating the encryption $l$ times, where each clone from the previous round is further cloned into $n+1$ new ones, the total number $N_{\text{clones}}$ of clones grows exponentially as $N_{\text{clones}}=(n+1)^{l+1}$. 

A notable feature of this scheme is that decrypting any single final clone does not require all intermediate noise qubits. Instead, the nested decryption procedure requires only $n$ noise qubits from each iteration level, amounting to a total of $n(l+1)$ noise qubits. This represents an exponential reduction in the decryption resources: the original encrypted cloning protocol, when producing the same number of clones, requires $(n+1)^{l+1}$ noise qubits for decryption.

As an experimental demonstration of iterated encrypted cloning, we implement the encrypted cloning protocol iteratively with $n=2$, yielding $3^{l+1}$ encrypted clones of qubit $A$, which is initially prepared in the maximally entangled state $\ket{\phi}$ with $\tilde{A}$. Since a IBM Heron R2 processor contains $156$ qubits, up to $l=2$ full iterations, requiring a total of $54$ qubits and producing $27$ encrypted clones, can be realized. However, by applying the encrypted cloning protocol to a subset of these $27$ clones, the number of encrypted clones can be further increased until all available qubits on the IBM Heron R2 chip are exhausted.

Table~\ref{tab:iterated_fidelity} (in Extended Data) shows the total number of qubits involved, the number of encrypted clones, the entanglement fidelity $F_e$, the number of 2-qubit gate containing layers $L_{2q}$ and the total number of 2-qubit gate operations $N_{2q}$ for $n=2$ encrypted cloning that is iterated $l$ times: 
For example, the experiment $l=0$, contains no iteration and produced $3$ encrypted clones, and decrypted one, while the $l=1$ experiment iterated once, i.e., it created three new clones for each clone, yielding a second generation of encrypted clones and decrypted one of them. Similarly, the $l=2$ experiment created a third generation of in total 27 clones and decrypted one of them. 
In the experiment labeled $3^\ddag$, we exhausted the Heron R2 chip's capacity of 156 qubits by creating as many fourth generation encrypted clones as possible. A full fourth generation would have yielded 81 encrypted clones with a total of 162 qubits. With the capacity of 156 qubits on the chip, we created a total of 77 encrypted clones of which 2 are of the third generation and 75 are of the fourth generation. We decrypted a clone of the fourth generation. 

We observe that, when iterating encrypted cloning with $n=2$ on the given hardware, we have an entanglement witness for up to $27$ encrypted clones and that the entanglement fidelity stays above the noise floor for up to 77 encrypted clones. 

As an intermediary experiment, labeled $l=2^*$, we also created 27 third generation qubits of which we then encrypted cloned 11 to obtain a total of 49 encrypted clones (16 of the third generation and 33 of the fourth generation). We decrypted one of the encrypted clones of the third generation. The entanglement fidelity drop from experiment $2$ to experiment $2^*$ demonstrates the deleterious effect of the increased time to decryption in experiment $2^*$ compared to experiment $2$.

We can read off that in $l$ times iterated cloning, entanglement can be witnessed, for $n=2$, for up to $l=2$, which produces $27$ encrypted clones. Let us compare this with non-iterated cloning. In comparison, in Experiment 1, we saw that the entanglement fidelity stays above the noise floor up to $n=11$, while we have an entanglement witness only up to $n=7$. 

We can conclude, therefore, that at least on the hardware that we used, large scale encrypted cloning is better performed iteratively, i.e., it is significantly more efficiently performed by raising $l$ than by raising $n$. 

In Supplementary Information, we describe a further investigation of iterating encrypted cloning, where we repeatedly cloned each time just one of the created clones. Given that the noise floor of the fidelity is at $0.25$, we there read off from the experimental results that in $l$ times iterated cloning a signal can be observed for $n=2$ up to $l=5$, which would produce $729$ encrypted clones.  

\medskip
\noindent
{\bf Experiment~4: Operation in parallel: Feasibility of encrypted cloning inside multipartite circuits}

\noindent
As a demonstration of the ability to create encrypted clones of genuine multipartite entanglement, we tested whether the protocol preserves the entanglement of $r$-partite GHZ state $\ket{\text{GHZ}_r} = (\ket{0}^{\otimes r} + \ket{1}^{\otimes r}) / \sqrt{2}$.
This family includes $\ket{+}=(\ket{0}+\ket{1})/\sqrt{2}$ and the Bell state $\ket{\phi}=(\ket{00}+\ket{11})/\sqrt{2}$ as the cases $r=1$ and $r=2$, respectively, while $r\geq 3$ corresponds to genuinely multipartite entangled states.

We prepared GHZ states for $r$ up to $15$. Each of the $r$ qubits is independently cloned using the $n=2$ encrypted cloning protocol, with its own distinct set of signal and noise qubits. This produces a system comprising $r$ independent groups of three encrypted clones. After the encrypted cloning, in each group, one of the clones (we chose $S_1$) is decrypted using all of the noise qubits in that group. Finally, the fidelity $F_r$ between the recovered GHZ state and the initial GHZ state is measured using the POM.

\medskip
\clearpage
\onecolumngrid
\noindent
{\bf \Large Extended Data}

\begin{table*}[h]
\caption{Measured entanglement fidelity $F_e$ of the reconstructed state and the number of 2-qubit gate
layers $L_\mathrm{2q}$ in Experiment~1 using either BSM or POM to measure the fidelity, as plotted in Fig.~\ref{fig:gate_depth_vs_n}. The table also lists the number of qubits $N_{\mathrm{qubits}}$ on the chip used for each case.}
\label{tab:EC_fidelity}
\begin{ruledtabular}
\begin{tabular}{cccccc}
\shortstack{Encrypted cloning \\ parameter $n$} & \shortstack{$N_\mathrm{qubits}$} & \shortstack{$L_\mathrm{2q}$ \\ BSM} & \shortstack{Fidelity $F_\mathrm{e}$ \\ BSM} &  \shortstack{$L_\mathrm{2q}$ \\ POM} & \shortstack{Fidelity $F_\mathrm{e}$ \\ POM} \\
\hline
2 & 6 & 21 & $0.823\pm 0.004$ & 18 & $0.875 \pm 0.008$ \\
3 & 8 & 33 & $0.839\pm 0.004$ & 25 & $0.810 \pm 0.008$\\
4 & 10 & 44 & $0.693\pm 0.005$ & 37 & $0.785 \pm 0.009$ \\
5 & 12 & 52 & $0.679\pm 0.005$ & 40 & $0.730 \pm 0.009$ \\
6 & 14 & 60 & $0.558\pm 0.005$ & 57 & $0.648 \pm 0.009$ \\
7 & 16 & 64 & $0.548\pm 0.005$ & 60 & $0.551 \pm 0.009$ \\
8 & 18 & 95 & $0.359\pm 0.005$ & 78 & $0.399 \pm 0.010$ \\
9 & 20 & 97 & $0.405\pm 0.005$ & 96 & $0.320 \pm 0.010$ \\
10 & 22 & 105 & $0.278 \pm 0.004$ & 107 & $0.351 \pm 0.010$ \\
11 & 24 & 115 & $0.327 \pm 0.005$ & 112 & $0.278 \pm 0.010$ \\
12 & 26 & 130 & $0.278 \pm 0.004$ & 101 & $0.316 \pm 0.010$ \\
13 & 28 & 143 & $0.289 \pm 0.005$ & 123 & $0.295 \pm 0.009$ \\
14 & 30 & 158 & $0.263 \pm 0.004$ & 155 & $0.250 \pm 0.009$ \\
15 & 32 & 152 & $0.270 \pm 0.004$ & 134 & $0.259 \pm 0.009$ \\
\end{tabular}
\end{ruledtabular}
\end{table*}

\begin{table*}[h!]
\caption{Measured CHSH parameter $S$ between the ancillary qubit $\tilde{A}$ and the decrypted encrypted clone in Experiment~2, as plotted in Fig.~\ref{fig:CHSH-b}.}
\label{tab:S_vs_n}
\begin{ruledtabular}
\begin{tabular}{c|ccc}
 & \multicolumn{3}{c}{CHSH parameter $S$} \\ 
\shortstack{$n$} & Scenario 2-1 & Scenario 2-2 & Scenario 2-3\\
\hline
2 & $ 2.147 \pm 0.017$ & $2.377 \pm 0.016$ & $2.331 \pm 0.016$ \\
3 & $ 1.695 \pm 0.018$ & $2.164 \pm 0.017$ & $2.087 \pm 0.017$ \\
4 & $ 1.474 \pm 0.019$ & $1.929 \pm 0.018$ & $1.901 \pm 0.018$ \\
5 & $ 0.703 \pm 0.020$ & $1.449 \pm 0.019$ & $1.293 \pm 0.019$ \\
6 & $ 0.432 \pm 0.020$ & $1.345 \pm 0.019$ & $1.200 \pm 0.019$ 
\end{tabular}
\end{ruledtabular}
\end{table*}

\begin{table*}[h]
\caption{Measured entanglement fidelities $F_e$ of the channels obtained from $l$ iterations of the $n=2$ encrypted cloning protocol and subsequent decryption in Experiment~3. The table also reports the number of qubits on the chip used in each case ($N_{\mathrm{qubits}}$), the number of two-qubit gate layers ($L_\mathrm{2q}$), and the number of two-qubit gates ($N_\mathrm{2q}$). In the case $l=2^*$, we created 16 third generation clones ($l=2$) plus 33 fourth generation clones, after which we decrypted a third generation clone. The result shows the deterioration of the fidelity in decrypting a third generation clone that is due to the extra wait time due to the creation of fourth generation clones. In the case $l=3^{\ddag}$, we generated fourth generation clones until the processor was filled and then decrypted one of these fourth generation encrypted clones.}
\label{tab:iterated_fidelity}
\begin{ruledtabular}
\begin{tabular}{cccccc}
$l$ & $N_{\mathrm{qubits}}$  & \makecell{Number of\\ clones} & $L_\mathrm{2q}$ &$N_\mathrm{2q}$ & Fidelity $F_e$ \\
\hline
0 & 6 & 3  & 18 & 21 & $0.875 \pm 0.008$ \\
1 & 18 & 9  & 38 & 81 & $0.723 \pm 0.009$ \\
2 & 54 & 27  & 63 & 299 & $0.569 \pm 0.009$\\
$2^*$ & 98 & 49 & 90 & 655  & $0.355 \pm 0.009$\\
$3^{\ddag}$ & 154 & 77 & 137 & 1422& $0.286 \pm 0.009$ 
\end{tabular}
\end{ruledtabular}
\end{table*}

\begin{table*}[t]
\caption{Measured fidelity $F_r$ of the reconstructed state with the original GHZ state in Experiment~4, as plotted in Fig.~\ref{fig:gate_depth_vs_r}. The table also reports the number of qubits on the chip used in each case ($N_{\mathrm{qubits}}$), the number of two-qubit gate layers ($L_\mathrm{2q}$), and the number of two-qubit gates ($N_\mathrm{2q}$). }
\label{tab:ghz_fidelity}
\begin{ruledtabular}
\begin{tabular}{ccccc}
\shortstack{Number of GHZ state qubits $r$} & \shortstack{$N_\mathrm{qubits}$} & $L_\mathrm{2q}$ & $N_\mathrm{2q}$ & \shortstack{Fidelity $F_r$} \\
\hline
1 & 5 & 19 & 21 & $0.928 \pm 0.009$ \\
2 & 10 & 20 & 41 & $0.786 \pm 0.009$ \\
3 & 15 & 25 & 71 & $0.709 \pm 0.008$ \\
4 & 20 & 27 & 98 & $0.575 \pm 0.008$ \\
5 & 25 & 36 & 128 & $0.451 \pm 0.008$ \\
6 & 30 & 42 & 163 & $0.353 \pm 0.007$ \\
7 & 35 & 49 & 198 & $0.322 \pm 0.007$ \\
8 & 40 & 54 & 225 & $0.258 \pm 0.006$ \\
9 & 45 & 69 & 248 & $0.193 \pm 0.006$ \\
10 & 50 & 63 & 312 & $0.170 \pm 0.006$ \\
11 & 55 & 73 & 350 & $0.134 \pm 0.005$ \\
12 & 60 & 81 & 341 & $0.123 \pm 0.005$ \\
13 & 65 & 87 & 368 & $0.065 \pm 0.004$ \\
14 & 70 & 95 & 481 & $0.049 \pm 0.004$ \\
15 & 75 & 116 & 519 & $0.039 \pm 0.004$ 
\end{tabular}
\end{ruledtabular}
\end{table*}

\newpage

\begin{figure*}[t] 
  \centering
  \begin{subfigure}{0.49\textwidth}
    \centering
    \includegraphics[width=\linewidth]{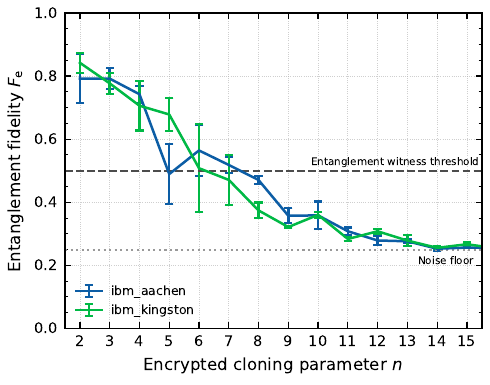}
    \caption{}
    \label{fig:aachen}
  \end{subfigure}\hfill
  \begin{subfigure}{0.49\textwidth}
    \centering
    \includegraphics[width=\linewidth]{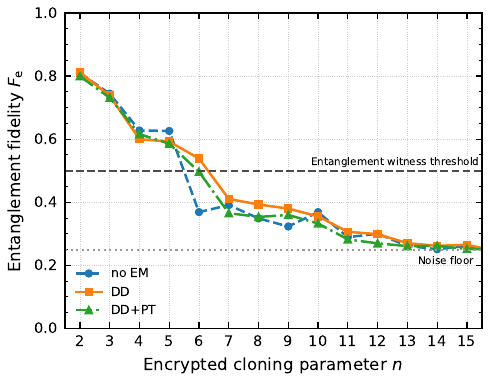}
    \caption{EM}
    \label{fig:EM}
  \end{subfigure}

  \caption{Entanglement fidelities $F_e$ of Experiment 1 under the influence of hardware variability and error mitigation methods (a) Comparison between multiple transpilation runs on \texttt{ibm\_kingston} and \texttt{ibm\_aachen} by averaging over 3 unique transpilation runs. Error bars are the standard deviation over the 3 runs. (b) Comparison between Experiment 1 using no error mitigation, dynamical decoupling and Pauli twirling in combination with dynamical decoupling.}
  \label{fig:comp_EM}
\end{figure*}

\newpage

\newpage

\clearpage
\twocolumngrid
\noindent
{\bf \Large Supplementary Information}

\medskip
\noindent
{\bf Simulation of experiments}

\noindent
While Experiments~1--4 were performed on the IBM Heron R2 processor \texttt{ibm\_kingston}, we also simulated all experiments that have a total qubit number of $\leq 27$ using Qiskit-Aer with the \texttt{statevector} method, in the absence of noise. In each case, we confirmed that encrypted cloning followed by decryption always yielded fidelity and entanglement fidelity values of 1. Simulating the Bell state experiment also yielded the theoretically expected CHSH parameter of $2\sqrt{2}$. The simulations using Qiskit-Aer, therefore, confirmed the theoretical result that the decryption of an encrypted clone perfectly recovers the quantum state that was encrypted cloned.  

\medskip
\noindent
{\bf Entanglement witness}

\noindent
An entanglement witness~\cite{guhne_entanglement_2009} is an observable $W$ that allows the detection of entanglement in a quantum state, satisfying $\mathrm{Tr}(W\rho_{\mathrm{sep}}) \geq 0$ for all separable states $\rho_{\mathrm{sep}}$, while for a given entangled state $\rho$, $\mathrm{Tr}(W\rho) < 0$.

In the case of the $r$-qubit GHZ state, a commonly used entanglement witness takes the form
\begin{equation}
    W_{\mathrm{GHZ}_r} = \frac{1}{2} I - |\mathrm{GHZ}_r\rangle\langle \mathrm{GHZ}_r|,
\end{equation}
where $I$ denotes the identity operator on the $2^r$-dimensional Hilbert space. This form of witness detects entanglement if $\mathrm{Tr}(W_{\mathrm{GHZ}_r}\rho) < 0$.

The expectation value of the witness can be expressed in terms of the fidelity $F_r = \langle \mathrm{GHZ}_r | \rho | \mathrm{GHZ}_r\rangle$ as $\mathrm{Tr}(W_{\mathrm{GHZ}_r}\rho) = \frac{1}{2} - F_r$.
Consequently, the condition for detecting entanglement becomes $F_r> \frac{1}{2}$. In experimental contexts, this relation provides a practical criterion for detecting GHZ-type multipartite entanglement by measuring the fidelity of the prepared state with respect to the ideal GHZ state.

\medskip
\noindent
{\bf Comparison with imperfect cloning}

\noindent
Theoretically, the encrypted cloning protocol enables perfect recovery of the original quantum state from any clone after decryption. In practice, however, the fidelity of the recovered state is degraded by experimental imperfections arising from the finite coherence time and gate infidelity of current quantum hardware. Despite these limitations, our experimental results demonstrate a clear advantage of the encrypted cloning protocol over the universal quantum cloning machine (UQCM)~\cite{buzek_quantum_1996}, which represents the theoretical limit of imperfect cloning.

The $1\to M$ UQCM~\cite{gisin_optimal_1997,werner_optimal_1998} is defined as
\begin{align}
    T(\rho)\coloneqq \frac{2}{M+1}\Pi_{\mathrm{sym}}^{(M)}\left(\rho\otimes \mathbb{I}^{\otimes (M-1)}\right)\Pi_{\mathrm{sym}}^{(M)},
\end{align}
where $I$ denotes the identity operator on a single-qubit system, and $\Pi_{\mathrm{sym}}^{(M)}$ represents the projection operator onto the symmetric subspace of the $M$-qubit system.

Define a channel $\Lambda$ by $\Lambda\coloneqq \mathrm{Tr}_{2,\cdots,M}\circ T$, which is obtained by tracing out $M-1$ qubits after applying $T$. It is known that the resulting channel is the depolarizing channel
\begin{align}
    \Lambda(\rho)=\eta \rho +(1-\eta)\frac{\mathbb{I}}{2}
\end{align}
with $\eta=\frac{M+2}{3M}$ (see, e.g., \cite{werner_optimal_1998}). 

Let us first calculate the entanglement fidelity of $\Lambda$. For $\ket{\phi}_{\tilde{A}A}=\frac{1}{\sqrt{2}}\left(\ket{0}_{\tilde{A}}\ket{0}_A+\ket{1}_{\tilde{A}}\ket{1}_A\right)$, the corresponding density operator, $\phi_{\tilde{A}A}\coloneqq \ket{\phi}\bra{\phi}_{\tilde{A}A}$ is expanded as
\begin{align}
    \phi_{\tilde{A}A}=\frac{1}{4}\left(\mathbb{I}\otimes\mathbb{I}+\sigma_1\otimes \sigma_1-\sigma_2\otimes \sigma_2+\sigma_3\otimes \sigma_1\right),
\end{align}
implying that 
\begin{align}
    &\mathcal{I}_{\tilde{A}}\otimes \Lambda (\phi_{\tilde{A}A})\nonumber\\
    &=\frac{1}{4}\left(\mathbb{I}\otimes\mathbb{I}+\eta\sigma_1\otimes \sigma_1-\eta\sigma_2\otimes \sigma_2+\eta\sigma_3\otimes \sigma_1\right).
\end{align}
Therefore, the entanglement fidelity of $\Lambda$ is given by
\begin{align}
    F_e\left(\phi,\mathcal{I}_{\tilde{A}}\otimes \Lambda (\phi_{\tilde{A}A})\right)=\frac{1}{4}(1+3\eta)=\frac{M+1}{2M}.
\end{align}
On the other hand, since $\Lambda^\dag (\sigma_i)=\eta\sigma_i$ holds for the dual channel $\Lambda^\dag$, the CHSH parameter for $\mathcal{I}_{\tilde{A}}\otimes \Lambda(\phi_{\tilde{A}A})$ is given by
\begin{align}
    S=\eta 2\sqrt{2}=\frac{M+2}{3M}2\sqrt{2},
\end{align}
where we have used the fact $S=2\sqrt{2}$ for $\phi_{\tilde{A}A}$. 

From the above calculations, for $M=3,5,7$, the entanglement fidelity is given by $F_e=0.667, 0.6, 0.571$, while the CHSH parameter is given by $S=1.57135, 1.31993, 1.21218$. We compare these theoretical values for UQCM with experimental results in Figs.~\ref{fig:gate_depth_vs_n} and \ref{fig:CHSH-b} for encrypted cloning with $n=2,3,4,5,6$, yielding $M=3,5,7$ clones, since $A$ also serves as an encrypted clone when $n$ is even. The observed entanglement fidelity of the encrypted cloning protocol in Experiment~1 exceeds the maximal theoretical value achievable by the UQCM for $n = 2, 3, 4,5$. Similarly, the experimentally observed CHSH parameter for encrypted cloning, at least in the Scenario~(2-3), is higher than the theoretical values for the UQCM for $n=2,3,4,5,6$. Notably, the encrypted cloning protocol experimentally violates the CHSH inequality, whereas the UQCM cannot do so even under ideal, error-free conditions. This finding highlights a fundamental distinction between the two approaches: while the UQCM inevitably degrades quantum correlations and cannot reproduce nonlocal quantum statistics, the encrypted cloning protocol fully preserves the underlying entanglement structure through encryption and subsequent decryption. 

We conclude with a comparison to the results of \cite{peng_cloning_2020}, which reported on an experimental implementation of the UQCM protocol, i.e., of imperfect cloning, for a Bell state in a photonic system, with two imperfect clones produced of each half of the Bell pair. There, the theoretical fidelity upper bounded by $0.583$ and the experimentally achieved fidelity of imperfectly cloned Bell state was $\approx 0.569 \pm 0.007$. In comparison, in our 
Experiment~4 with $r=2$, we start with a Bell state and create 3 encrypted clones of each half of the Bell pair. After decrypting one of each triplet of encrypted clones, we measure for the resulting pair decrypted clones the fidelity to the original Bell pair. While the theoretical fidelity is of course 1, the experimentally achieved fidelity in our experiments was $0.786 \pm 0.009$, which is well above even the theoretical upper bound of the UQCM protocol.

\medskip
\noindent
{\bf Relativistic Considerations for Delayed-Choice Experiments}

\noindent
Experiment~2 introduces a novel framework for a delayed-choice experiment. After preparing the entangled pair $(A, \tilde{A})$ and creating $n$ encrypted clones of qubit $A$, the qubit $\tilde{A}$ is no longer maximally entangled with any individual qubit nor with $A$ itself.

Instead, $\tilde{A}$ possesses delocalized entanglement across the entire system of clones and the decryption key.
The subsequent measurement of $\tilde{A}$ then nonlocally updates the global state of the system of encrypted clones and the key, without a locally observable effect. 

By choosing and then decrypting an encrypted clone $S_j$, the entanglement of $\tilde{A}$ becomes re-localized in the sense that the chosen $S_j$ becomes the Bell partner qubit of $\tilde{A}$, whose state was collapsed by the measurement of $\tilde{A}$. This demonstrates a novel kind of control over where and when the consequences of a local measurement within a spatially spread-out entangled system appear.

Alternatively, if a clone $S_j$ is decrypted and measured first, that measurement then collapses the state of the distant ancilla $\tilde{A}$. This reversed measurement ordering highlights that, following encryption, we can choose which clone to decrypt and by decrypting it we ensure that exclusively its measurement will then non-locally collapse the state of $\tilde{A}$. This demonstrates a new way to effectively re-localize entanglement among multiple potential endpoints.

Our experiment in the Scenarios~2-1, 2-2 and 2-3 proved that  
pre-existing Bell state entanglement between qubits $\tilde{A}$ and $A$ is indeed preserved through the de-localizing and then re-localizing encryption-decryption channel and, therefore, is accessible in the controllable, delayed-choice manner discussed above.

In particular, by measuring correlations in these delayed-choice setups, we tested the violation of the CHSH inequality between a decrypted clone and the ancilla $\tilde{A}$, in order to confirm the recovery of genuine quantumness in the decryption process. 

In principle, in future experiments with a sufficiently distant ancilla $\tilde{A}$, the measurements on $\tilde{A}$ and an encrypted clone could be performed spacelike separated, so that the very same scenario is then a case of 2-1, 2-2 or 2-3, dependent on the observer's state of motion. 

This scenario then generalizes a well-known phenomenon of quantum entanglement: when measurements of two qubits $A$ and $\tilde{A}$ in a Bell state are performed in a spacelike separated way, then the observer's state of motion determines whether the measurement outcome of either $A$ or $\tilde{A}$ is predetermined. This is because for an observer where the measurement of $A$ occurs before $\tilde{A}$, $A$'s measurement determines $\tilde{A}$'s outcome, making it predetermined due to state collapse, or vice versa for an observer in whose frame $\tilde{A}$ is measured first.

The key novelty now is that, in addition to choosing an observer, encrypted cloning introduces a second choice: which encrypted clone of $A$ to decrypt. 
Therefore, bringing encrypted cloning into this scenario allows one to choose which is the partner qubit to $\tilde{A}$ that ends up in a Bell state.

Crucially, this choice can also be spacelike separated from the measurement of $\tilde{A}$, leading to a delayed choice phenomenon. For an observer who measures $\tilde{A}$ first, 
the choice of which encrypted clone to decrypt (and thus which clone is the partner to $\tilde{A}$) may not have been made yet, i.e., can be made after $\tilde{A}$ is measured, thereby making a delayed choice of which specific clone will be the one whose state was collapsed by the prior measurement of $\tilde{A}$. 

Vice versa, we can consider observers for whom the choice of which encrypted clone, $S_j$, is being decrypted and measured occurs before the measurement of $\tilde{A}$. For these observers, the choice of which $S_j$ to decrypt determines which signal qubit has the power to collapse the state of $\tilde{A}$ when measured.  

Introducing encrypted cloning, therefore, offers an intriguing new perspective on the interplay between causality and quantum non-locality, showing that, in the above sense, the distribution of entanglement over subsystems can be subject to an observer-dependent and also delayed-choice-dependent selection process.

\medskip
\noindent
{\bf Confirmation of the mixedness of encrypted clone}

\begin{figure*}[t] 
  \centering
  \begin{subfigure}{0.49\textwidth}
    \centering
    \includegraphics[width=\linewidth]{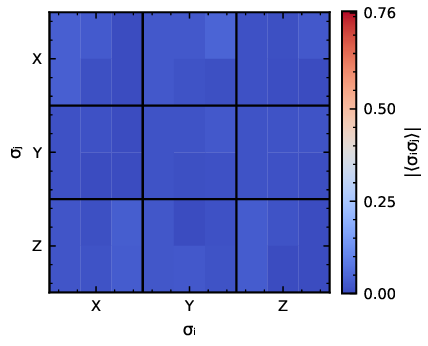}
    \caption{}
    \label{fig:sub1}
  \end{subfigure}\hfill
  \begin{subfigure}{0.49\textwidth}
    \centering
    \includegraphics[width=\linewidth]{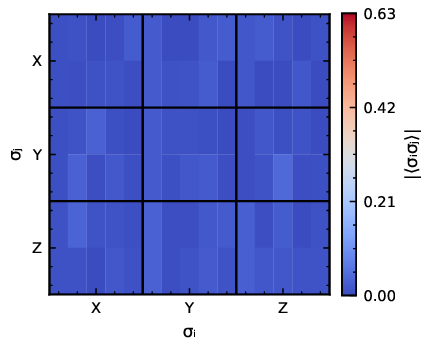}
    \caption{}
    \label{fig:sub3}
  \end{subfigure}

  \vspace{0.6em}

  \begin{subfigure}{0.49\textwidth}
    \centering
    \includegraphics[width=\linewidth]{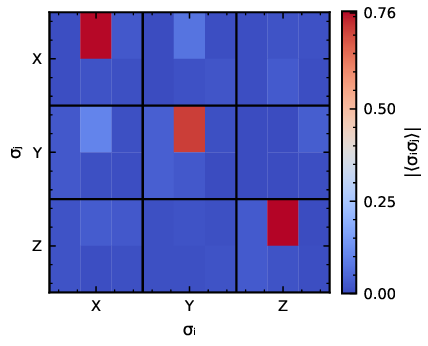}
    \caption{}
    \label{fig:sub4}
  \end{subfigure}\hfill
  \begin{subfigure}{0.49\textwidth}
    \centering
    \includegraphics[width=\linewidth]{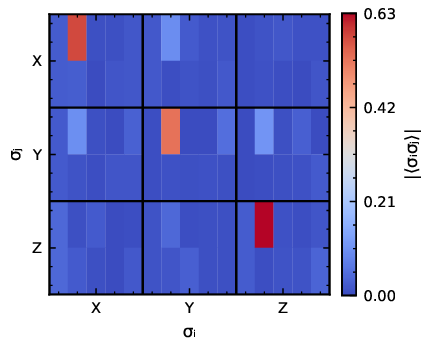}
    \caption{}
    \label{fig:sub6}
  \end{subfigure}

  \caption{Correlation strength of the two-point correlator $T_{ij}$ between $\Tilde{A}$ and all other qubits. Each block is separated into a left sub-block: qubit $A$, top row: signal qubits $S_i$, bottom row: noise qubits $N_i$. (a),(b) Correlation strength after encryption for encrypted cloning with $n=2,\, 4$. (c),(d): Correlation strength after encryption and subsequent decryption for encrypted cloning with $n=2,\, 4$.}
  \label{fig:Correlations}
\end{figure*}

\noindent
In order to verify that after encryption, the clones and noise qubits are in a fully mixed state and no information about the entanglement with $\Tilde{A}$ can be gathered, we measure the correlations between $\Tilde{A}$ and all other qubits involved in the cloning protocol. To this end, we experimentally determine the full two-point correlation matrix $T \in \mathbb{R}^{3\times 3}$ with 
\begin{equation}
    T_{ij} = \mathrm{Tr}[\rho_\mathrm{exp}(\sigma_i \otimes \sigma_j)], \: i,j = 1,2,3 \: ,
\end{equation}
being a measurement of the expectation value on $\Tilde{A}$ and $A$, $S_k$ or $N_k$.
Figure \ref{fig:Correlations} show the correlation strength $\abs{T_{ij}}$ for encrypted cloning with $n=2,4$ before and after decryption on the IBM device \texttt{ibm\_kingston}, using the Estimator primitive from Qiskit IBM Runtime, with default settings to measure all $T_{ij}$.

The results are consistent with the theoretical prediction that the clones and noise qubits are not individually correlated to $\Tilde{A}$.
In addition, after decoding the correlation strength with $\Tilde{A}$ is mostly located on the decoded qubit and diagonal of the correlation matrix $T$. With growing system size, hardware errors leak correlation strength to other qubits and decrease the fidelity of the reconstructed state.

\medskip
\noindent
{\bf Hardware variability}

\noindent
We analyzed how different transpilation runs and quantum chips impact the entanglement fidelity in Experiment~1.
In Fig.~\ref{fig:comp_EM}, we executed Experiment~1 on \texttt{ibm\_aachen} and \texttt{ibm\_kingston} with identical transpiler settings, with only the transpiler seed being different over a total of 3 runs. It can be observed that the fidelity depends strongly on the physical execution of the quantum circuit, with the experiments on \texttt{ibm\_aachen} producing a higher entanglement fidelity for most executions.

Also, to account for any variability in the performance of the quantum hardware over time, experimental trials were conducted across five distinct sessions spanning several weeks on \texttt{ibm\_kingston}. The measured fidelity $F_\mathrm{e}$ of the $n=2$ case in Experiment 1 exhibited fluctuations reflecting the impact of daily recalibration cycles and varying transpilation on the device. The mean fidelity over five runs is $F_\mathrm{e} = 0.833 \pm 0.027$ (see also Table~\ref{tab:hardware_volatility}). This uncertainty interval represents the standard deviation across sessions, providing a conservative estimate of the system's performance stability under recalibration of the quantum hardware. 

\begin{table}[]
    \centering
    \caption{Measured entanglement fidelities $F_\mathrm{e}$ of $n=2$ encrypted cloning in Experiment~1 using POM, with measurements taken on different days.}
    \begin{tabular}{c |c}
        Date (YYYY-MM-DD) & Fidelity $F_\mathrm{e}$ \\
        \hline
        2025-11-18 & $0.824 \pm 0.008$\\
        2025-11-20 & $0.875 \pm 0.008$\\
        2025-12-06 & $0.810 \pm 0.009$\\
        2026-01-16 & $0.844 \pm 0.009$\\
        2026-01-18 & $0.813 \pm 0.008$ \\ 
    \end{tabular}
    \label{tab:hardware_volatility}
\end{table}

\medskip
\noindent
{\bf Error mitigation and suppression}

\noindent
We performed Experiments~1 and~3 using the following two quantum error suppression techniques:
\begin{itemize}
    \item Dynamical Decoupling (DD) mitigates qubit decoherence and crosstalk by inserting sequences of gates (that cumulatively act as the logical identity) during idle periods of qubit evolution (see, e.g, \cite{PhysRevApplied.20.064027} for the theory behind DD).
    DD is particularly useful when certain qubits have long idle times during the circuit execution. Dynamical decoupling is implemented within the Sampler primitive of the Qiskit IBM Runtime. The pulse sequence applied is the \textit{XY}4-sequence.
    
    \item Pauli Twirling (PT) transforms arbitrary noise channels into stochastic, Pauli noise channels by sandwiching layers of two-qubit gates by Pauli gates that are sampled at random, such that the logical operation of these layers is preserved. The goal of PT is to mitigate coherent noise. This is desirable because coherent noise grows quadratically with the depth of a quantum circuit, in contrast to Pauli noise which only grows linearly. Like DD, PT is implemented within the Sampler primitive in Qiskit IBM Runtime. We used the default Sampler settings which turns on measurement and gate twirling using $10 000$ shots, 313 shots per randomization and 32 randomizations.
\end{itemize}
More details on the theory and implementation of each of these techniques in Qiskit IBM Runtime may be found in \cite{emdocs}. \\

\begin{figure}[htbp]
  \centering
  \includegraphics[width=\linewidth]{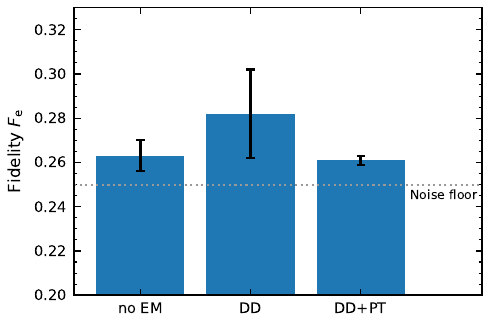} 
  \caption{Fidelity of iterated encrypted cloning in the case $l = 3^{\ddag}$ using no error mitigation (no EM), dynamical decoupling (DD) and Pauli twirling in combination with dynamical decoupling (DD + PT). Values are averaged over 3 different transpilation runs. Errors are the standard deviation over the 3 runs.}
  \label{fig:FullQ}
\end{figure}

Applied to the standard encrypted cloning (Experiment~1) and the iterated encrypted cloning filling the maximum \texttt{ibm\_kingston} chip ($l=3^{\ddag}$ in Experiment~3), we found that they do not improve the performance of our experiments significantly, see Figs.~\ref{fig:comp_EM} and~\ref{fig:FullQ}.

\medskip
\noindent
{\bf Maximally iterated cloning}

\begin{figure}
    \centering
    \includegraphics{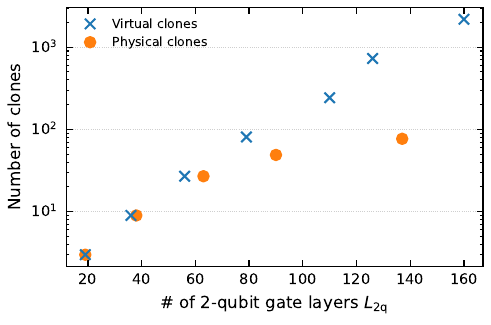}
    \caption{In Experiment 3, a logarithmic plot of the number of encrypted clones created by iterated encrypted cloning versus the circuit depth. The number of physically created encrypted clones grows sub-exponentially with the circuit depth as the chip fills up quickly and the compilations become constrained and less efficient. When encrypted cloning only one clone per iteration (the others remaining `virtual'), the chip fills up more slowly and the exponential increase persists longer.}
    \label{fig:vir_phys_number_depth}
\end{figure}

\begin{figure}
    \centering
    \includegraphics{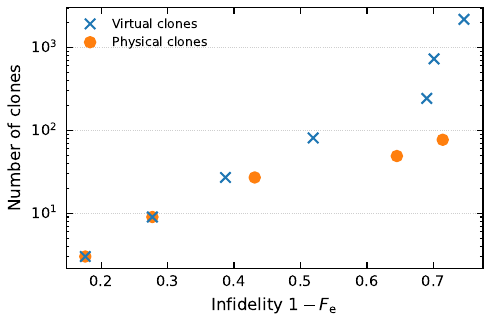}
    \caption{In Experiment 3, the number of physical, i.e., of actually produced encrypted clones and the number of `virtual' encrypted clones, i.e., the number of encrypted clones producible if the chip contained more qubits, versus the infidelity of the decryption. In the latter case, in each iteration step, we create clones of only one clone. This leaves more space on the chip, allowing compilations with lower circuit depths and, as shown in the plot, with correspondingly lower infidelities.}
    \label{fig:vir_phys_number_infidelity}
\end{figure}

\begin{figure}
    \centering
    \includegraphics{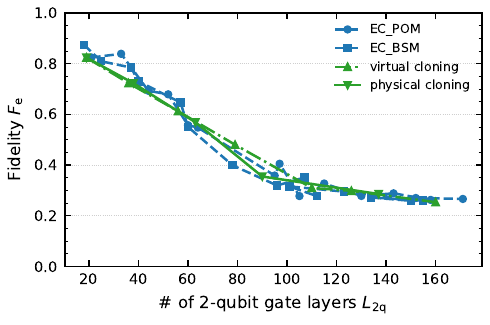}
    \caption{In Experiments 1 and 3, a plot of the observed entanglement fidelities versus the two-qubit layer depth.
    The plots ``EC\_POM'' and ``EC\_BSM'' show the results for Experiment~1 with POM and BSM, respectively. 
    The plots ``virtual cloning'' and ``physical cloning'' show the results for Experiment~3; in the former, we repeatedly cloned each time just one of the created clones; in the latter, all the clones of each generation are encrypted cloned in each iteration. The close similarity of these plots shows that the degradation of the fidelity of the decrypted clone is governed by the circuit depth. }
    \label{fig:vir_phys_exp1_fid_depth}
\end{figure}

\noindent
We here report the experimental result of an extension of Experiment~3, i.e., iterating encrypted cloning: In practice, iterated cloning is limited by both noise and chip size. In order to explore the maximal experimental repeatability of encrypted cloning irrespective of chip size, we repeatedly cloned in each iteration only one of the created clones. Table~\ref{tab:iterated} shows the observed fidelities. We read off that the result is above the noise floor (i.e., $F_e>1/4$) for $n=2$ up to $l=5$. This means that if, with the same performance, the chip had sufficiently many qubits, then $729$ encrypted clones could be produced that are above the noise floor. For more details, see also Figs.~\ref{fig:vir_phys_number_depth}, \ref{fig:vir_phys_number_infidelity}, and \ref{fig:vir_phys_exp1_fid_depth}.

\begin{table*}[t]
\caption{Measured entanglement fidelities $F_e$ of the channel obtained via $l$ iterations of the $n=2,4$ encrypted cloning and subsequent decryption of a signal qubit when only a single signal qubit is re-cloned. The number of qubits required in decryption grows linearly with $l$, while the number of virtual clones $N_{\mathrm{virt. clones}}=(n+1)^{(l+1)}$ grows exponentially. Here, ``virtual clones'' refers to the encrypted clones that would be produced if the encrypted cloning protocol were applied to all clones in every iteration. 
Given that the threshold for an entanglement witness is a fidelity of $1/2$, the table shows that with the hardware we used, it is possible to create $N_{\mathrm{virt. clones}}=27$ encrypted clones so that each encrypted clone, if decrypted, possesses provable entanglement with the ancilla $\tilde{A}$. The table also reports the number of qubits $N_{\mathrm{qubits}}$ on the chip used for each case, the number of two-qubit gate layers $L_\mathrm{2q}$, and the number of two-qubit gates $N_\mathrm{2q}$.}
\begin{ruledtabular}\label{tab:iterated}
\begin{tabular}{c|ccccc|ccccc}
& \multicolumn{5}{c|}{Encrypted cloning $n=2$} & \multicolumn{5}{c}{Encrypted cloning $n=4$}\\
Iteration $l$ & $N_\mathrm{qubits}$ & $N_\mathrm{virt. clones}$ & $L_\mathrm{2q}$ & $N_\mathrm{2q}$ & $F_e$ & $N_\mathrm{qubits}$ & $N_\mathrm{virt. clones}$ & $L_\mathrm{2q}$ & $N_\mathrm{2q}$ & $F_e$\\
\hline
0 & 6  & 3    & 18  & 21  & $0.875 \pm 0.008$ & 10 & 5     & 44  & 57  & $0.693 \pm 0.009$\\
1 & 10 & 9    & 36  & 44  & $0.723 \pm 0.009$ & 18 & 25    & 83  & 142 & $0.450 \pm 0.010$\\
2 & 14 & 27   & 56  & 79  & $0.613 \pm 0.009$ & 26 & 125   & 145 & 222 & $0.291 \pm 0.009$\\
3 & 18 & 81   & 79  & 121 & $0.481 \pm 0.010$ & 34 & 625   & 176 & 310 & $0.255 \pm 0.009$\\
4 & 22 & 243  & 110 & 167 & $0.310 \pm 0.009$  &    &       &     &     &  \\
5 & 26 & 729  & 126 & 197 & $0.299 \pm 0.009$ &    &       &     &     &  \\
6 & 30 & 2187 & 160 & 226 & $0.254 \pm 0.009$ &    &       &     &     &   
\end{tabular}
\end{ruledtabular}
\end{table*}

\begin{thebibliography}{33}%
\makeatletter
\providecommand \@ifxundefined [1]{%
 \@ifx{#1\undefined}
}%
\providecommand \@ifnum [1]{%
 \ifnum #1\expandafter \@firstoftwo
 \else \expandafter \@secondoftwo
 \fi
}%
\providecommand \@ifx [1]{%
 \ifx #1\expandafter \@firstoftwo
 \else \expandafter \@secondoftwo
 \fi
}%
\providecommand \natexlab [1]{#1}%
\providecommand \enquote  [1]{``#1''}%
\providecommand \bibnamefont  [1]{#1}%
\providecommand \bibfnamefont [1]{#1}%
\providecommand \citenamefont [1]{#1}%
\providecommand \href@noop [0]{\@secondoftwo}%
\providecommand \href [0]{\begingroup \@sanitize@url \@href}%
\providecommand \@href[1]{\@@startlink{#1}\@@href}%
\providecommand \@@href[1]{\endgroup#1\@@endlink}%
\providecommand \@sanitize@url [0]{\catcode `\\12\catcode `\$12\catcode `\&12\catcode `\#12\catcode `\^12\catcode `\_12\catcode `\%12\relax}%
\providecommand \@@startlink[1]{}%
\providecommand \@@endlink[0]{}%
\providecommand \url  [0]{\begingroup\@sanitize@url \@url }%
\providecommand \@url [1]{\endgroup\@href {#1}{\urlprefix }}%
\providecommand \urlprefix  [0]{URL }%
\providecommand \Eprint [0]{\href }%
\providecommand \doibase [0]{https://doi.org/}%
\providecommand \selectlanguage [0]{\@gobble}%
\providecommand \bibinfo  [0]{\@secondoftwo}%
\providecommand \bibfield  [0]{\@secondoftwo}%
\providecommand \translation [1]{[#1]}%
\providecommand \BibitemOpen [0]{}%
\providecommand \bibitemStop [0]{}%
\providecommand \bibitemNoStop [0]{.\EOS\space}%
\providecommand \EOS [0]{\spacefactor3000\relax}%
\providecommand \BibitemShut  [1]{\csname bibitem#1\endcsname}%
\let\auto@bib@innerbib\@empty
\bibitem [{\citenamefont {Wootters}\ and\ \citenamefont {Zurek}(1982)}]{wootters_single_1982}%
  \BibitemOpen
  \bibfield  {author} {\bibinfo {author} {\bibfnamefont {W.~K.}\ \bibnamefont {Wootters}}\ and\ \bibinfo {author} {\bibfnamefont {W.~H.}\ \bibnamefont {Zurek}},\ }\href {https://doi.org/10.1038/299802a0} {\bibfield  {journal} {\bibinfo  {journal} {Nature}\ }\textbf {\bibinfo {volume} {299}},\ \bibinfo {pages} {802} (\bibinfo {year} {1982})}\BibitemShut {NoStop}%
\bibitem [{\citenamefont {Dieks}(1982)}]{dieks_communication_1982}%
  \BibitemOpen
  \bibfield  {author} {\bibinfo {author} {\bibfnamefont {D.}~\bibnamefont {Dieks}},\ }\href {https://doi.org/10.1016/0375-9601(82)90084-6} {\bibfield  {journal} {\bibinfo  {journal} {Physics Letters A}\ }\textbf {\bibinfo {volume} {92}},\ \bibinfo {pages} {271} (\bibinfo {year} {1982})}\BibitemShut {NoStop}%
\bibitem [{\citenamefont {Bužek}\ and\ \citenamefont {Hillery}(1996)}]{buzek_quantum_1996}%
  \BibitemOpen
  \bibfield  {author} {\bibinfo {author} {\bibfnamefont {V.}~\bibnamefont {Bužek}}\ and\ \bibinfo {author} {\bibfnamefont {M.}~\bibnamefont {Hillery}},\ }\href {https://doi.org/10.1103/PhysRevA.54.1844} {\bibfield  {journal} {\bibinfo  {journal} {Physical Review A}\ }\textbf {\bibinfo {volume} {54}},\ \bibinfo {pages} {1844} (\bibinfo {year} {1996})}\BibitemShut {NoStop}%
\bibitem [{\citenamefont {Gisin}\ and\ \citenamefont {Massar}(1997)}]{gisin_optimal_1997}%
  \BibitemOpen
  \bibfield  {author} {\bibinfo {author} {\bibfnamefont {N.}~\bibnamefont {Gisin}}\ and\ \bibinfo {author} {\bibfnamefont {S.}~\bibnamefont {Massar}},\ }\href {https://doi.org/10.1103/PhysRevLett.79.2153} {\bibfield  {journal} {\bibinfo  {journal} {Physical Review Letters}\ }\textbf {\bibinfo {volume} {79}},\ \bibinfo {pages} {2153} (\bibinfo {year} {1997})}\BibitemShut {NoStop}%
\bibitem [{\citenamefont {Bruß}\ \emph {et~al.}(1998)\citenamefont {Bruß}, \citenamefont {DiVincenzo}, \citenamefont {Ekert}, \citenamefont {Fuchs}, \citenamefont {Macchiavello},\ and\ \citenamefont {Smolin}}]{brus_optimal_1998}%
  \BibitemOpen
  \bibfield  {author} {\bibinfo {author} {\bibfnamefont {D.}~\bibnamefont {Bruß}}, \bibinfo {author} {\bibfnamefont {D.~P.}\ \bibnamefont {DiVincenzo}}, \bibinfo {author} {\bibfnamefont {A.}~\bibnamefont {Ekert}}, \bibinfo {author} {\bibfnamefont {C.~A.}\ \bibnamefont {Fuchs}}, \bibinfo {author} {\bibfnamefont {C.}~\bibnamefont {Macchiavello}},\ and\ \bibinfo {author} {\bibfnamefont {J.~A.}\ \bibnamefont {Smolin}},\ }\href {https://doi.org/10.1103/PhysRevA.57.2368} {\bibfield  {journal} {\bibinfo  {journal} {Physical Review A}\ }\textbf {\bibinfo {volume} {57}},\ \bibinfo {pages} {2368} (\bibinfo {year} {1998})}\BibitemShut {NoStop}%
\bibitem [{\citenamefont {Werner}(1998)}]{werner_optimal_1998}%
  \BibitemOpen
  \bibfield  {author} {\bibinfo {author} {\bibfnamefont {R.~F.}\ \bibnamefont {Werner}},\ }\href {https://doi.org/10.1103/PhysRevA.58.1827} {\bibfield  {journal} {\bibinfo  {journal} {Physical Review A}\ }\textbf {\bibinfo {volume} {58}},\ \bibinfo {pages} {1827} (\bibinfo {year} {1998})}\BibitemShut {NoStop}%
\bibitem [{\citenamefont {Duan}\ and\ \citenamefont {Guo}(1998)}]{duan_probabilistic_1998}%
  \BibitemOpen
  \bibfield  {author} {\bibinfo {author} {\bibfnamefont {L.-M.}\ \bibnamefont {Duan}}\ and\ \bibinfo {author} {\bibfnamefont {G.-C.}\ \bibnamefont {Guo}},\ }\href {https://doi.org/10.1103/PhysRevLett.80.4999} {\bibfield  {journal} {\bibinfo  {journal} {Physical Review Letters}\ }\textbf {\bibinfo {volume} {80}},\ \bibinfo {pages} {4999} (\bibinfo {year} {1998})}\BibitemShut {NoStop}%
\bibitem [{\citenamefont {Pati}(1999)}]{pati_quantum_1999}%
  \BibitemOpen
  \bibfield  {author} {\bibinfo {author} {\bibfnamefont {A.~K.}\ \bibnamefont {Pati}},\ }\href {https://doi.org/10.1103/PhysRevLett.83.2849} {\bibfield  {journal} {\bibinfo  {journal} {Physical Review Letters}\ }\textbf {\bibinfo {volume} {83}},\ \bibinfo {pages} {2849} (\bibinfo {year} {1999})}\BibitemShut {NoStop}%
\bibitem [{\citenamefont {Fan}\ \emph {et~al.}(2001)\citenamefont {Fan}, \citenamefont {Matsumoto}, \citenamefont {Wang},\ and\ \citenamefont {Wadati}}]{fan_quantum_2001}%
  \BibitemOpen
  \bibfield  {author} {\bibinfo {author} {\bibfnamefont {H.}~\bibnamefont {Fan}}, \bibinfo {author} {\bibfnamefont {K.}~\bibnamefont {Matsumoto}}, \bibinfo {author} {\bibfnamefont {X.-B.}\ \bibnamefont {Wang}},\ and\ \bibinfo {author} {\bibfnamefont {M.}~\bibnamefont {Wadati}},\ }\href {https://doi.org/10.1103/PhysRevA.65.012304} {\bibfield  {journal} {\bibinfo  {journal} {Physical Review A}\ }\textbf {\bibinfo {volume} {65}},\ \bibinfo {pages} {012304} (\bibinfo {year} {2001})}\BibitemShut {NoStop}%
\bibitem [{\citenamefont {Wang}\ \emph {et~al.}(2011)\citenamefont {Wang}, \citenamefont {Shi}, \citenamefont {Xiong}, \citenamefont {Jing}, \citenamefont {Ren}, \citenamefont {Mu},\ and\ \citenamefont {Fan}}]{wang_unified_2011}%
  \BibitemOpen
  \bibfield  {author} {\bibinfo {author} {\bibfnamefont {Y.-N.}\ \bibnamefont {Wang}}, \bibinfo {author} {\bibfnamefont {H.-D.}\ \bibnamefont {Shi}}, \bibinfo {author} {\bibfnamefont {Z.-X.}\ \bibnamefont {Xiong}}, \bibinfo {author} {\bibfnamefont {L.}~\bibnamefont {Jing}}, \bibinfo {author} {\bibfnamefont {X.-J.}\ \bibnamefont {Ren}}, \bibinfo {author} {\bibfnamefont {L.-Z.}\ \bibnamefont {Mu}},\ and\ \bibinfo {author} {\bibfnamefont {H.}~\bibnamefont {Fan}},\ }\href {https://doi.org/10.1103/PhysRevA.84.034302} {\bibfield  {journal} {\bibinfo  {journal} {Physical Review A}\ }\textbf {\bibinfo {volume} {84}},\ \bibinfo {pages} {034302} (\bibinfo {year} {2011})}\BibitemShut {NoStop}%
\bibitem [{\citenamefont {Bell}(1964)}]{bell_einstein_1964}%
  \BibitemOpen
  \bibfield  {author} {\bibinfo {author} {\bibfnamefont {J.~S.}\ \bibnamefont {Bell}},\ }\href {https://doi.org/10.1103/PhysicsPhysiqueFizika.1.195} {\bibfield  {journal} {\bibinfo  {journal} {Physics Physique Fizika}\ }\textbf {\bibinfo {volume} {1}},\ \bibinfo {pages} {195} (\bibinfo {year} {1964})}\BibitemShut {NoStop}%
\bibitem [{\citenamefont {Clauser}\ \emph {et~al.}(1969)\citenamefont {Clauser}, \citenamefont {Horne}, \citenamefont {Shimony},\ and\ \citenamefont {Holt}}]{clauser_proposed_1969}%
  \BibitemOpen
  \bibfield  {author} {\bibinfo {author} {\bibfnamefont {J.~F.}\ \bibnamefont {Clauser}}, \bibinfo {author} {\bibfnamefont {M.~A.}\ \bibnamefont {Horne}}, \bibinfo {author} {\bibfnamefont {A.}~\bibnamefont {Shimony}},\ and\ \bibinfo {author} {\bibfnamefont {R.~A.}\ \bibnamefont {Holt}},\ }\href {https://doi.org/10.1103/PhysRevLett.23.880} {\bibfield  {journal} {\bibinfo  {journal} {Physical Review Letters}\ }\textbf {\bibinfo {volume} {23}},\ \bibinfo {pages} {880} (\bibinfo {year} {1969})}\BibitemShut {NoStop}%
\bibitem [{\citenamefont {Scarani}\ \emph {et~al.}(2005)\citenamefont {Scarani}, \citenamefont {Iblisdir}, \citenamefont {Gisin},\ and\ \citenamefont {Acín}}]{scarani_quantum_2005}%
  \BibitemOpen
  \bibfield  {author} {\bibinfo {author} {\bibfnamefont {V.}~\bibnamefont {Scarani}}, \bibinfo {author} {\bibfnamefont {S.}~\bibnamefont {Iblisdir}}, \bibinfo {author} {\bibfnamefont {N.}~\bibnamefont {Gisin}},\ and\ \bibinfo {author} {\bibfnamefont {A.}~\bibnamefont {Acín}},\ }\href {https://doi.org/10.1103/RevModPhys.77.1225} {\bibfield  {journal} {\bibinfo  {journal} {Reviews of Modern Physics}\ }\textbf {\bibinfo {volume} {77}},\ \bibinfo {pages} {1225} (\bibinfo {year} {2005})}\BibitemShut {NoStop}%
\bibitem [{\citenamefont {Fan}\ \emph {et~al.}(2014)\citenamefont {Fan}, \citenamefont {Wang}, \citenamefont {Jing}, \citenamefont {Yue}, \citenamefont {Shi}, \citenamefont {Zhang},\ and\ \citenamefont {Mu}}]{fan_quantum_2014}%
  \BibitemOpen
  \bibfield  {author} {\bibinfo {author} {\bibfnamefont {H.}~\bibnamefont {Fan}}, \bibinfo {author} {\bibfnamefont {Y.-N.}\ \bibnamefont {Wang}}, \bibinfo {author} {\bibfnamefont {L.}~\bibnamefont {Jing}}, \bibinfo {author} {\bibfnamefont {J.-D.}\ \bibnamefont {Yue}}, \bibinfo {author} {\bibfnamefont {H.-D.}\ \bibnamefont {Shi}}, \bibinfo {author} {\bibfnamefont {Y.-L.}\ \bibnamefont {Zhang}},\ and\ \bibinfo {author} {\bibfnamefont {L.-Z.}\ \bibnamefont {Mu}},\ }\href {https://doi.org/10.1016/j.physrep.2014.06.004} {\bibfield  {journal} {\bibinfo  {journal} {Physics Reports}\ }\bibinfo {series} {Quantum cloning machines and the applications},\ \textbf {\bibinfo {volume} {544}},\ \bibinfo {pages} {241} (\bibinfo {year} {2014})}\BibitemShut {NoStop}%
\bibitem [{\citenamefont {Lamas-Linares}\ \emph {et~al.}(2002)\citenamefont {Lamas-Linares}, \citenamefont {Simon}, \citenamefont {Howell},\ and\ \citenamefont {Bouwmeester}}]{lamas-linares_experimental_2002}%
  \BibitemOpen
  \bibfield  {author} {\bibinfo {author} {\bibfnamefont {A.}~\bibnamefont {Lamas-Linares}}, \bibinfo {author} {\bibfnamefont {C.}~\bibnamefont {Simon}}, \bibinfo {author} {\bibfnamefont {J.~C.}\ \bibnamefont {Howell}},\ and\ \bibinfo {author} {\bibfnamefont {D.}~\bibnamefont {Bouwmeester}},\ }\href {https://doi.org/10.1126/science.1068972} {\bibfield  {journal} {\bibinfo  {journal} {Science}\ }\textbf {\bibinfo {volume} {296}},\ \bibinfo {pages} {712} (\bibinfo {year} {2002})}\BibitemShut {NoStop}%
\bibitem [{\citenamefont {Sciarrino}\ and\ \citenamefont {De~Martini}(2005)}]{sciarrino_realization_2005}%
  \BibitemOpen
  \bibfield  {author} {\bibinfo {author} {\bibfnamefont {F.}~\bibnamefont {Sciarrino}}\ and\ \bibinfo {author} {\bibfnamefont {F.}~\bibnamefont {De~Martini}},\ }\href {https://doi.org/10.1103/PhysRevA.72.062313} {\bibfield  {journal} {\bibinfo  {journal} {Physical Review A}\ }\textbf {\bibinfo {volume} {72}},\ \bibinfo {pages} {062313} (\bibinfo {year} {2005})}\BibitemShut {NoStop}%
\bibitem [{\citenamefont {Du}\ \emph {et~al.}(2005)\citenamefont {Du}, \citenamefont {Durt}, \citenamefont {Zou}, \citenamefont {Li}, \citenamefont {Kwek}, \citenamefont {Lai}, \citenamefont {Oh},\ and\ \citenamefont {Ekert}}]{du_experimental_2005}%
  \BibitemOpen
  \bibfield  {author} {\bibinfo {author} {\bibfnamefont {J.}~\bibnamefont {Du}}, \bibinfo {author} {\bibfnamefont {T.}~\bibnamefont {Durt}}, \bibinfo {author} {\bibfnamefont {P.}~\bibnamefont {Zou}}, \bibinfo {author} {\bibfnamefont {H.}~\bibnamefont {Li}}, \bibinfo {author} {\bibfnamefont {L.~C.}\ \bibnamefont {Kwek}}, \bibinfo {author} {\bibfnamefont {C.~H.}\ \bibnamefont {Lai}}, \bibinfo {author} {\bibfnamefont {C.~H.}\ \bibnamefont {Oh}},\ and\ \bibinfo {author} {\bibfnamefont {A.}~\bibnamefont {Ekert}},\ }\href {https://doi.org/10.1103/PhysRevLett.94.040505} {\bibfield  {journal} {\bibinfo  {journal} {Physical Review Letters}\ }\textbf {\bibinfo {volume} {94}},\ \bibinfo {pages} {040505} (\bibinfo {year} {2005})}\BibitemShut {NoStop}%
\bibitem [{\citenamefont {Chen}\ \emph {et~al.}(2007)\citenamefont {Chen}, \citenamefont {Zhou}, \citenamefont {Suter},\ and\ \citenamefont {Du}}]{chen_experimental_2007}%
  \BibitemOpen
  \bibfield  {author} {\bibinfo {author} {\bibfnamefont {H.}~\bibnamefont {Chen}}, \bibinfo {author} {\bibfnamefont {X.}~\bibnamefont {Zhou}}, \bibinfo {author} {\bibfnamefont {D.}~\bibnamefont {Suter}},\ and\ \bibinfo {author} {\bibfnamefont {J.}~\bibnamefont {Du}},\ }\href {https://doi.org/10.1103/PhysRevA.75.012317} {\bibfield  {journal} {\bibinfo  {journal} {Physical Review A}\ }\textbf {\bibinfo {volume} {75}},\ \bibinfo {pages} {012317} (\bibinfo {year} {2007})}\BibitemShut {NoStop}%
\bibitem [{\citenamefont {Nagali}\ \emph {et~al.}(2009)\citenamefont {Nagali}, \citenamefont {Sansoni}, \citenamefont {Sciarrino}, \citenamefont {De~Martini}, \citenamefont {Marrucci}, \citenamefont {Piccirillo}, \citenamefont {Karimi},\ and\ \citenamefont {Santamato}}]{nagali_optimal_2009}%
  \BibitemOpen
  \bibfield  {author} {\bibinfo {author} {\bibfnamefont {E.}~\bibnamefont {Nagali}}, \bibinfo {author} {\bibfnamefont {L.}~\bibnamefont {Sansoni}}, \bibinfo {author} {\bibfnamefont {F.}~\bibnamefont {Sciarrino}}, \bibinfo {author} {\bibfnamefont {F.}~\bibnamefont {De~Martini}}, \bibinfo {author} {\bibfnamefont {L.}~\bibnamefont {Marrucci}}, \bibinfo {author} {\bibfnamefont {B.}~\bibnamefont {Piccirillo}}, \bibinfo {author} {\bibfnamefont {E.}~\bibnamefont {Karimi}},\ and\ \bibinfo {author} {\bibfnamefont {E.}~\bibnamefont {Santamato}},\ }\href {https://doi.org/10.1038/nphoton.2009.214} {\bibfield  {journal} {\bibinfo  {journal} {Nature Photonics}\ }\textbf {\bibinfo {volume} {3}},\ \bibinfo {pages} {720} (\bibinfo {year} {2009})}\BibitemShut {NoStop}%
\bibitem [{\citenamefont {Pan}\ \emph {et~al.}(2011)\citenamefont {Pan}, \citenamefont {Liu}, \citenamefont {Yang},\ and\ \citenamefont {Fan}}]{pan_solid-state_2011}%
  \BibitemOpen
  \bibfield  {author} {\bibinfo {author} {\bibfnamefont {X.-Y.}\ \bibnamefont {Pan}}, \bibinfo {author} {\bibfnamefont {G.-Q.}\ \bibnamefont {Liu}}, \bibinfo {author} {\bibfnamefont {L.-L.}\ \bibnamefont {Yang}},\ and\ \bibinfo {author} {\bibfnamefont {H.}~\bibnamefont {Fan}},\ }\href {https://doi.org/10.1063/1.3624595} {\bibfield  {journal} {\bibinfo  {journal} {Applied Physics Letters}\ }\textbf {\bibinfo {volume} {99}},\ \bibinfo {pages} {051113} (\bibinfo {year} {2011})}\BibitemShut {NoStop}%
\bibitem [{\citenamefont {Peng}\ \emph {et~al.}(2020)\citenamefont {Peng}, \citenamefont {Wu}, \citenamefont {Zhong}, \citenamefont {Luo}, \citenamefont {Li}, \citenamefont {Hu}, \citenamefont {Jiang}, \citenamefont {Chen}, \citenamefont {Li}, \citenamefont {Liu}, \citenamefont {Nemoto}, \citenamefont {Munro}, \citenamefont {Sanders}, \citenamefont {Lu},\ and\ \citenamefont {Pan}}]{peng_cloning_2020}%
  \BibitemOpen
  \bibfield  {author} {\bibinfo {author} {\bibfnamefont {L.-C.}\ \bibnamefont {Peng}}, \bibinfo {author} {\bibfnamefont {D.}~\bibnamefont {Wu}}, \bibinfo {author} {\bibfnamefont {H.-S.}\ \bibnamefont {Zhong}}, \bibinfo {author} {\bibfnamefont {Y.-H.}\ \bibnamefont {Luo}}, \bibinfo {author} {\bibfnamefont {Y.}~\bibnamefont {Li}}, \bibinfo {author} {\bibfnamefont {Y.}~\bibnamefont {Hu}}, \bibinfo {author} {\bibfnamefont {X.}~\bibnamefont {Jiang}}, \bibinfo {author} {\bibfnamefont {M.-C.}\ \bibnamefont {Chen}}, \bibinfo {author} {\bibfnamefont {L.}~\bibnamefont {Li}}, \bibinfo {author} {\bibfnamefont {N.-L.}\ \bibnamefont {Liu}}, \bibinfo {author} {\bibfnamefont {K.}~\bibnamefont {Nemoto}}, \bibinfo {author} {\bibfnamefont {W.~J.}\ \bibnamefont {Munro}}, \bibinfo {author} {\bibfnamefont {B.~C.}\ \bibnamefont {Sanders}}, \bibinfo {author} {\bibfnamefont {C.-Y.}\ \bibnamefont {Lu}},\ and\ \bibinfo {author} {\bibfnamefont {J.-W.}\ \bibnamefont {Pan}},\ }\href {https://doi.org/10.1103/PhysRevLett.125.210502}
  {\bibfield  {journal} {\bibinfo  {journal} {Physical Review Letters}\ }\textbf {\bibinfo {volume} {125}},\ \bibinfo {pages} {210502} (\bibinfo {year} {2020})}\BibitemShut {NoStop}%
\bibitem [{\citenamefont {Yang}\ \emph {et~al.}(2021)\citenamefont {Yang}, \citenamefont {Han}, \citenamefont {Huang}, \citenamefont {Ning}, \citenamefont {Li}, \citenamefont {Xu}, \citenamefont {Zheng}, \citenamefont {Fan},\ and\ \citenamefont {Zheng}}]{yang_experimental_2021}%
  \BibitemOpen
  \bibfield  {author} {\bibinfo {author} {\bibfnamefont {Z.-B.}\ \bibnamefont {Yang}}, \bibinfo {author} {\bibfnamefont {P.-R.}\ \bibnamefont {Han}}, \bibinfo {author} {\bibfnamefont {X.-J.}\ \bibnamefont {Huang}}, \bibinfo {author} {\bibfnamefont {W.}~\bibnamefont {Ning}}, \bibinfo {author} {\bibfnamefont {H.}~\bibnamefont {Li}}, \bibinfo {author} {\bibfnamefont {K.}~\bibnamefont {Xu}}, \bibinfo {author} {\bibfnamefont {D.}~\bibnamefont {Zheng}}, \bibinfo {author} {\bibfnamefont {H.}~\bibnamefont {Fan}},\ and\ \bibinfo {author} {\bibfnamefont {S.-B.}\ \bibnamefont {Zheng}},\ }\href {https://doi.org/10.1038/s41534-021-00375-5} {\bibfield  {journal} {\bibinfo  {journal} {npj Quantum Information}\ }\textbf {\bibinfo {volume} {7}},\ \bibinfo {pages} {44} (\bibinfo {year} {2021})}\BibitemShut {NoStop}%
\bibitem [{\citenamefont {Yamaguchi}\ and\ \citenamefont {Kempf}(2026)}]{yamaguchi_encrypted_2026}%
  \BibitemOpen
  \bibfield  {author} {\bibinfo {author} {\bibfnamefont {K.}~\bibnamefont {Yamaguchi}}\ and\ \bibinfo {author} {\bibfnamefont {A.}~\bibnamefont {Kempf}},\ }\href {https://doi.org/10.1103/y4y1-1ll6} {\bibfield  {journal} {\bibinfo  {journal} {Physical Review Letters}\ }\textbf {\bibinfo {volume} {136}},\ \bibinfo {pages} {010801} (\bibinfo {year} {2026})}\BibitemShut {NoStop}%
\bibitem [{\citenamefont {Schumacher}(1996)}]{schumacher_sending_1996}%
  \BibitemOpen
  \bibfield  {author} {\bibinfo {author} {\bibfnamefont {B.}~\bibnamefont {Schumacher}},\ }\href {https://doi.org/10.1103/PhysRevA.54.2614} {\bibfield  {journal} {\bibinfo  {journal} {Physical Review A}\ }\textbf {\bibinfo {volume} {54}},\ \bibinfo {pages} {2614} (\bibinfo {year} {1996})}\BibitemShut {NoStop}%
\bibitem [{\citenamefont {Horodecki}\ \emph {et~al.}(1999)\citenamefont {Horodecki}, \citenamefont {Horodecki},\ and\ \citenamefont {Horodecki}}]{horodecki_general_1999}%
  \BibitemOpen
  \bibfield  {author} {\bibinfo {author} {\bibfnamefont {M.}~\bibnamefont {Horodecki}}, \bibinfo {author} {\bibfnamefont {P.}~\bibnamefont {Horodecki}},\ and\ \bibinfo {author} {\bibfnamefont {R.}~\bibnamefont {Horodecki}},\ }\href {https://doi.org/10.1103/PhysRevA.60.1888} {\bibfield  {journal} {\bibinfo  {journal} {Physical Review A}\ }\textbf {\bibinfo {volume} {60}},\ \bibinfo {pages} {1888} (\bibinfo {year} {1999})}\BibitemShut {NoStop}%
\bibitem [{\citenamefont {Nielsen}(2002)}]{nielsen_simple_2002}%
  \BibitemOpen
  \bibfield  {author} {\bibinfo {author} {\bibfnamefont {M.~A.}\ \bibnamefont {Nielsen}},\ }\href {https://doi.org/10.1016/S0375-9601(02)01272-0} {\bibfield  {journal} {\bibinfo  {journal} {Physics Letters A}\ }\textbf {\bibinfo {volume} {303}},\ \bibinfo {pages} {249} (\bibinfo {year} {2002})}\BibitemShut {NoStop}%
\bibitem [{\citenamefont {Greenberger}\ \emph {et~al.}(1990)\citenamefont {Greenberger}, \citenamefont {Horne}, \citenamefont {Shimony},\ and\ \citenamefont {Zeilinger}}]{greenberger_bells_1990}%
  \BibitemOpen
  \bibfield  {author} {\bibinfo {author} {\bibfnamefont {D.~M.}\ \bibnamefont {Greenberger}}, \bibinfo {author} {\bibfnamefont {M.~A.}\ \bibnamefont {Horne}}, \bibinfo {author} {\bibfnamefont {A.}~\bibnamefont {Shimony}},\ and\ \bibinfo {author} {\bibfnamefont {A.}~\bibnamefont {Zeilinger}},\ }\href {https://doi.org/10.1119/1.16243} {\bibfield  {journal} {\bibinfo  {journal} {American Journal of Physics}\ }\textbf {\bibinfo {volume} {58}},\ \bibinfo {pages} {1131} (\bibinfo {year} {1990})}\BibitemShut {NoStop}%
\bibitem [{\citenamefont {Javadi-Abhari}\ \emph {et~al.}(2024)\citenamefont {Javadi-Abhari}, \citenamefont {Treinish}, \citenamefont {Krsulich}, \citenamefont {Wood}, \citenamefont {Lishman}, \citenamefont {Gacon}, \citenamefont {Martiel}, \citenamefont {Nation}, \citenamefont {Bishop}, \citenamefont {Cross}, \citenamefont {Johnson},\ and\ \citenamefont {Gambetta}}]{qiskit2024}%
  \BibitemOpen
  \bibfield  {author} {\bibinfo {author} {\bibfnamefont {A.}~\bibnamefont {Javadi-Abhari}}, \bibinfo {author} {\bibfnamefont {M.}~\bibnamefont {Treinish}}, \bibinfo {author} {\bibfnamefont {K.}~\bibnamefont {Krsulich}}, \bibinfo {author} {\bibfnamefont {C.~J.}\ \bibnamefont {Wood}}, \bibinfo {author} {\bibfnamefont {J.}~\bibnamefont {Lishman}}, \bibinfo {author} {\bibfnamefont {J.}~\bibnamefont {Gacon}}, \bibinfo {author} {\bibfnamefont {S.}~\bibnamefont {Martiel}}, \bibinfo {author} {\bibfnamefont {P.~D.}\ \bibnamefont {Nation}}, \bibinfo {author} {\bibfnamefont {L.~S.}\ \bibnamefont {Bishop}}, \bibinfo {author} {\bibfnamefont {A.~W.}\ \bibnamefont {Cross}}, \bibinfo {author} {\bibfnamefont {B.~R.}\ \bibnamefont {Johnson}},\ and\ \bibinfo {author} {\bibfnamefont {J.~M.}\ \bibnamefont {Gambetta}},\ }\href {https://doi.org/10.48550/arXiv.2405.08810} {\bibinfo {title} {Quantum computing with {Q}iskit}} (\bibinfo {year} {2024}),\ \Eprint {https://arxiv.org/abs/2405.08810} {arXiv:2405.08810 [quant-ph]}
  \BibitemShut {NoStop}%
\bibitem [{\citenamefont {Tóth}\ and\ \citenamefont {Gühne}(2005)}]{toth_detecting_2005}%
  \BibitemOpen
  \bibfield  {author} {\bibinfo {author} {\bibfnamefont {G.}~\bibnamefont {Tóth}}\ and\ \bibinfo {author} {\bibfnamefont {O.}~\bibnamefont {Gühne}},\ }\href {https://doi.org/10.1103/PhysRevLett.94.060501} {\bibfield  {journal} {\bibinfo  {journal} {Physical Review Letters}\ }\textbf {\bibinfo {volume} {94}},\ \bibinfo {pages} {060501} (\bibinfo {year} {2005})}\BibitemShut {NoStop}%
\bibitem [{\citenamefont {Gühne}\ \emph {et~al.}(2007)\citenamefont {Gühne}, \citenamefont {Lu}, \citenamefont {Gao},\ and\ \citenamefont {Pan}}]{guhne_toolbox_2007}%
  \BibitemOpen
  \bibfield  {author} {\bibinfo {author} {\bibfnamefont {O.}~\bibnamefont {Gühne}}, \bibinfo {author} {\bibfnamefont {C.-Y.}\ \bibnamefont {Lu}}, \bibinfo {author} {\bibfnamefont {W.-B.}\ \bibnamefont {Gao}},\ and\ \bibinfo {author} {\bibfnamefont {J.-W.}\ \bibnamefont {Pan}},\ }\href {https://doi.org/10.1103/PhysRevA.76.030305} {\bibfield  {journal} {\bibinfo  {journal} {Physical Review A}\ }\textbf {\bibinfo {volume} {76}},\ \bibinfo {pages} {030305} (\bibinfo {year} {2007})}\BibitemShut {NoStop}%
\bibitem [{\citenamefont {Gühne}\ and\ \citenamefont {Tóth}(2009)}]{guhne_entanglement_2009}%
  \BibitemOpen
  \bibfield  {author} {\bibinfo {author} {\bibfnamefont {O.}~\bibnamefont {Gühne}}\ and\ \bibinfo {author} {\bibfnamefont {G.}~\bibnamefont {Tóth}},\ }\href {https://doi.org/10.1016/j.physrep.2009.02.004} {\bibfield  {journal} {\bibinfo  {journal} {Physics Reports}\ }\textbf {\bibinfo {volume} {474}},\ \bibinfo {pages} {1} (\bibinfo {year} {2009})}\BibitemShut {NoStop}%
\bibitem [{\citenamefont {Ezzell}\ \emph {et~al.}(2023)\citenamefont {Ezzell}, \citenamefont {Pokharel}, \citenamefont {Tewala}, \citenamefont {Quiroz},\ and\ \citenamefont {Lidar}}]{PhysRevApplied.20.064027}%
  \BibitemOpen
  \bibfield  {author} {\bibinfo {author} {\bibfnamefont {N.}~\bibnamefont {Ezzell}}, \bibinfo {author} {\bibfnamefont {B.}~\bibnamefont {Pokharel}}, \bibinfo {author} {\bibfnamefont {L.}~\bibnamefont {Tewala}}, \bibinfo {author} {\bibfnamefont {G.}~\bibnamefont {Quiroz}},\ and\ \bibinfo {author} {\bibfnamefont {D.~A.}\ \bibnamefont {Lidar}},\ }\href {https://doi.org/10.1103/PhysRevApplied.20.064027} {\bibfield  {journal} {\bibinfo  {journal} {Phys. Rev. Appl.}\ }\textbf {\bibinfo {volume} {20}},\ \bibinfo {pages} {064027} (\bibinfo {year} {2023})}\BibitemShut {NoStop}%
\bibitem [{emd(2025)}]{emdocs}%
  \BibitemOpen
  \href@noop {} {\bibinfo {title} {{Error mitigation and suppression techniques}}},\ \bibinfo {howpublished} {\url{https://quantum.cloud.ibm.com/docs/en/guides/error-mitigation-and-suppression-techniques}} (\bibinfo {year} {2025}),\ \bibinfo {note} {accessed: October 29, 2025}\BibitemShut {NoStop}%
\end{thebibliography}
\end{document}